\documentclass[onecolumn,preprintnumbers,amsmath,amssymb,11pt]{revtex4}
\usepackage{amsmath}
\usepackage{rotating}
\usepackage{amsmath}
\usepackage{extarrows}
\usepackage{bbm}
\usepackage[ruled]{algorithm2e}
\usepackage{graphicx}
\usepackage{dcolumn}
\usepackage{bm}
\usepackage[all]{xy}
\usepackage{indentfirst}
\usepackage{amsmath}
\usepackage{multirow}
\usepackage{mathrsfs}
\usepackage{euscript}
\usepackage{amssymb}
\begin{document}
\title{Computationally Efficient Nonlinear Bell Inequalities for Quantum Networks}
 \author{Ming-Xing Luo}

\affiliation{\small $^1$Information Security and National Computing Grid Laboratory,
\\
\small Southwest Jiaotong University, Chengdu 610031, China
\\
\small $^2$CSNMT, International Cooperation Research Center of China, Chengdu 610031, China
\\
\small $^3$Department of Physics, University of Michigan, Ann Arbor, MI 48109, USA}

\begin{abstract}
The correlations in quantum networks have attracted strong interest with new types of violations of the locality. The standard Bell inequalities cannot characterize the multipartite correlations that are generated by multiple sources. The main problem is that no computationally efficient method is available for constructing useful Bell inequalities for general quantum networks. In this work, we show a significant improvement by presenting new, explicit Bell-type inequalities for general networks including cyclic networks. These nonlinear inequalities are related to the matching problem of an equivalent unweighted bipartite graph that allows constructing a polynomial-time algorithm. For the quantum resources consisting of bipartite entangled pure states and generalized Greenberger-Horne-Zeilinger (GHZ) states, we prove the generic non-multilocality of quantum networks with multiple independent observers using new Bell inequalities. The violations are maximal with respect to the presented Tsirelson's bound for Einstein-Podolsky-Rosen (EPR) states and GHZ states. Moreover, these violations hold for Werner states or some general noisy states. Our results suggest that the presented Bell inequalities can be used to characterize experimental quantum networks.
\end{abstract}
\maketitle

\section*{Introduction}

Bell's well-known theorem \cite{Bel1} states that the predictions of quantum mechanics are inconsistent with classical causal relations that originate from a common local hidden variable (LHV). Specifically, the correlation between the outcomes of local measurements on a remotely shared entangled state cannot be described by a locally causal model. The study of quantum nonlocality has stimulated both remarkable developments in quantum theory \cite{Bel2,CB,MY,BCPS} and potential applications \cite{Ek,ABGM,PAM,BCMD,BGK}.

Quantum nonlocality has been significantly generalized by considering complex causal structures beyond the standard LHV models \cite{Pop,BGP,Fri1,HLP,GWCA,CMG,Fri2}. These improvements aim to provide rigorous theoretical frameworks of causal relations and structures \cite{CB,Pear,ABHL,SGS} and are useful for deriving Bell inequalities \cite{Bel2,CB,MY,BCPS,BCMD,RBG}. These inequalities are applicable for those networks with a single source. Nonetheless, for general networks there are several independent sources for distributing hidden states to space-like separated parties in terms of generalized locally causal model (GLCM) \cite{CB,Pear,ABHL,SGS}. As reasonable extensions of a single source, the multipartite correlations should be defined by multiple sources. Meanwhile, a meaningful Bell-type inequality enables the characterization of these correlations across the entire network. How to feature and verify the nonlocality of multipartite correlations not only are theoretically important to prove the supremacy \cite{BGK}, but also are experimentally challenging in the implementation of quantum networks \cite{ACL,Kim} and quantum repeaters \cite{SSdG}.

Unfortunately, the linear Bell inequalities derived from one source are useless for characterizing the multipartite correlations of general quantum networks. Recently, for the simplest network of entanglement swapping, nonlinear Bell inequalities have been proposed to verify the non-bilocality of tripartite correlations \cite{BGP,BRGP,TSCA,BBBC}. It is then extended for verifying the non-multilocality of general star-shaped networks \cite{ACSC}. For small-sized networks, the computational algebraic method \cite{LS} and linear programming technique \cite{Cha} provide reasonable routes to construct polynomial Bell inequalities. Another interesting method is to iteratively expand a given network to the desired network by adding independent sources \cite{RBBP}. Despite these advances, no computationally efficient method is available to feature general quantum networks. Additionally, the nonlinear Bell inequalities imply that some projection subspaces of the multipartite correlation space are not convex \cite{BRGP,LS,Cha,RBBP,GMTR}, which reveal new features beyond the correlation polytopes bounded by linear Bell inequalities \cite{Bel1,Bel2,CB,MY,BCPS}. A natural problem is whether these characteristics are typical for quantum networks. One of our goals is to address this problem. The nonlocality of some quantum networks have been experimentally verified using different physical systems \cite{SBBP,CASB,ACSB,HZHL}.

In this work, we propose simple and efficient nonlinear Bell inequalities to characterize the multipartite correlations of general quantum networks in terms of the GLCM \cite{Fri2,Pear,ABHL,SGS}. Notably, our approach depends primarily on the maximal matching problem of the equivalent unweighted bipartite graph \cite{West}, which allows constructing new Bell inequalities within polynomial time complexity. We further prove that the multipartite quantum correlations violate the presented nonlinear inequalities for all finite-size quantum networks with multiple observers that do not share entangled states. This violation or non-multilocality holds for the quantum resources consisting of all bipartite entangled pure states and generalized Greenberger-Horne-Zeilinger (GHZ) states, and can be maximal with respect to the Tsirelson's bound. The generic non-multilocality is different from the nonlocality of a single entangled pair using linear Bell inequality \cite{CFS,Gis} or CHSH inequality \cite{PR,CHSH}. Finally, we evaluate the upper bound of the critical visibilities of Werner states and general noisy states for which the non-multilocality is also true \cite{Cha,RBBP,GMTR}. Remarkably, our result holds for lots of cyclic networks that have not been investigated \cite{SSdG,BBBC,ACSC,LS,Cha,RBBP,GMTR}. The simplicity of our Bell inequalities makes them useful for experimental quantum networks.

\section*{Results}

\textit{Multilocality structure of a network.} In what follows, we consider the simplest scenario of dichotomic inputs and outputs for all parties.

Inspired from Bell inequalities of two parties \cite{Bel1}, the multilocality of correlations of a network follows from the GLCM \cite{CB,Pear,ABHL,SGS}. Formally, all systems measured in the experiment are considered to be in the hidden states of $\Lambda=(\lambda_1, \lambda_2, \cdots, \lambda_m)$, where $\Lambda$ are arbitrary and could exist prior to the measurement choices, and $m$ is the number of hidden states. The dichotomic output $a_i$ of any particular system can arbitrarily depend on hidden states $\Lambda$ and the type of measurement but not on the measurements performed on systems (here, one bit $x_j$ denotes the type of measurement). Thus, the GLCM suggests a joint conditional probability distribution of the measurement outcomes as
\begin{eqnarray}
P({\bf a}|{\bf x})=\int_{\Omega} d\mu(\Lambda) \prod_{i=1}^nP(a_i|x_i,\Lambda),
\label{eqn-1}
\end{eqnarray}
where $\textbf{a}=(a_1,a_2, \cdots, a_n)$, $\textbf{x}=(x_1, x_2, \cdots, x_n)$, $a_i, x_i\in \{0, 1\}$, and $(\Omega, \Sigma, \mu)$ denotes the measure space of hidden states $\Lambda$. In Eq.(\ref{eqn-1}), $\mu(\Lambda)$ is the measure of $\Lambda$ with the normalization condition $\int_\Omega d\mu(\Lambda)=1$, $P(a_i|x_i,\Lambda)$ is the conditional probability of the outcome $a_i$ for the $i$-th party (with the knowledge of $x_i$ and $\Lambda$) and satisfies $\sum_{a_i}P(a_i|x_i,\Lambda)=1$ for each $x_i$ and $\Lambda$, and $n$ is the number of space-like separated parties.

\begin{figure}
\begin{center}
\resizebox{200pt}{155pt}{\includegraphics{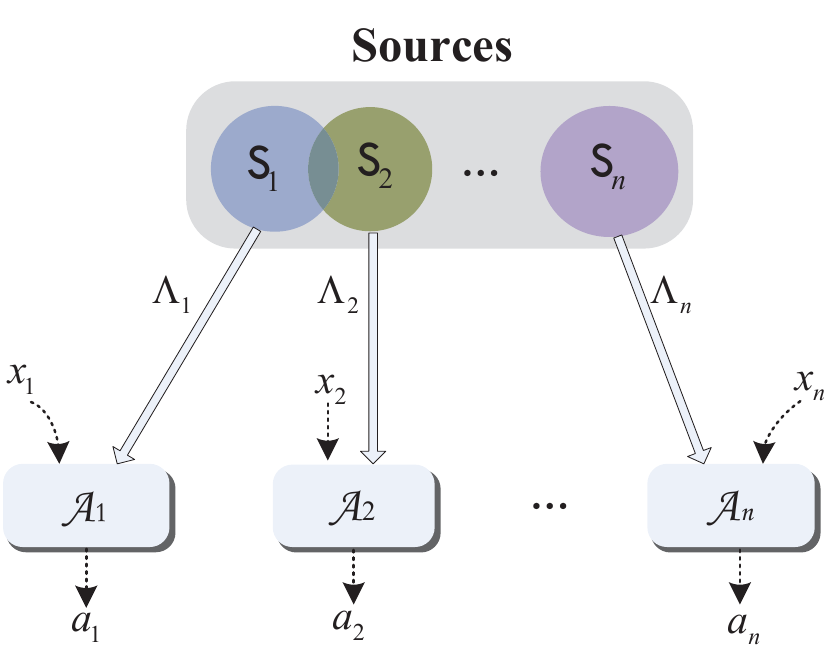}}
\end{center}
\caption{\small (Color online) Schematic network in terms of the GLCM. There are $m$ independent sources $S_1, S_2, \cdots, S_{m}$ that distribute hidden states $\lambda_1, \lambda_2, \cdots, \lambda_{m}$, respectively. Each space-like separated party ${\cal A}_i$ receives hidden states $\Lambda_{i}=\{\lambda_{j_1}, \lambda_{j_2}, \cdots, \lambda_{j_{\ell_i}}\}$ that are distributed by the corresponding sources $\textsf{S}_i=\{S_{j_1}, S_{j_2}, \cdots, S_{j_{\ell_i}}\}$, where $\cup_{i=1}^n\textsf{{S}}_i=\{S_1, S_2, \cdots, S_{m}\}$. In the experiment, each party ${\cal A}_i$ obtains one bit $a_i$ dependent on  input bit $x_i$ and hidden states $\Lambda_{i}$, $i=1, 2, \cdots, n$. One vector connects a source set and one party (cubic).}
\label{fig-1}
\end{figure}

Now, we consider a finite-size network shown in Fig.1 in terms of the GLCM. Assume that $m$ independent sources $S_1, S_2, \cdots, S_{m}$ distribute the  hidden states $\lambda_1, \lambda_2, \cdots, \lambda_{m}$, respectively. Each space-like separated party ${\cal A}_i$ receives hidden states $\Lambda_{i}=\{\lambda_{j_{1}}, \lambda_{j_2}, \cdots, \lambda_{j_{\ell_i}}\}$ from the corresponding sources $\textsf{{S}}_i=\{S_{j_{1}}, S_{j_2}, \cdots, S_{j_{\ell_i}}\}$, where $\textsf{{S}}_1, \cdots, \textsf{{S}}_n$ satisfy $\cup_{i=1}^n\textsf{{S}}_i=\{S_1, S_2, \cdots, S_{m}\}$. The measure of hidden states is given by $\mu(\Lambda)=\prod_{i=1}^m\mu_i(\lambda_i)$, where $\mu_i(\lambda_i)$ is the measure of $\lambda_i$ with the normalization condition $\int_{\Omega_i} d\mu_i(\lambda_i)=1$, and $(\Omega_i, \Sigma_i, \mu_i)$ denotes the measure space of $\lambda_i$, $i=1, 2, \cdots, m$. Eq.(\ref{eqn-1}) can be rewritten as
\begin{eqnarray}
P({\bf a}|{\bf x})
=\int_{\Omega} \prod_{i=1}^md\mu_i(\lambda_i)\prod_{j=1}^{n} P(a_j|x_j,\Lambda_{j}).
\label{eqn-2}
\end{eqnarray}
In the case of $m=1$, Eq.(\ref{eqn-2}) reduces to the locality assumption of one source and geometrically defines a correlation polytope, which contains all LHV distributions inside with the linear Bell inequalities as its facets \cite{Bel1,Pear,ABHL,SGS}. Unfortunately, similar correlation polytope does not exist for networks with multiple sources. Especially, the statistical correlations of the standard entanglement swapping \cite{BGP} imply a non-convex set consisting of tripartite correlations \cite{SSdG}. For some cases of $m>2$, new correlation sets may be elucidated by exploring nonlinear Bell-type inequalities according to the acyclic graph approach \cite{GM,LS}, linear programming technique \cite{Cha,Tar,KT} or network expansion \cite{RBBP}.

\textit{Explicit nonlinear Bell inequalities for networks.} Our method is based on geometric features of networks. A network is called {\it $k$-independent} if there are $k$ space-like separated parties that do not share sources. The $k$-independence is equivalent to the following {\it $k$-locality} in terms of the GLCM: there are $k$ subsets $\Lambda_{i_1}, \Lambda_{i_2}, \cdots, \Lambda_{i_k}$ consisting of hidden states such that
\begin{eqnarray}
\left\{
\begin{array}{ll}
\cup_{j=1}^k\Lambda_{i_j}\subseteq \Lambda,
\\
\Lambda_{i_s}\cap \Lambda_{i_t}=\varnothing \mbox { for all $s, t$ satisfying $1\leq s <t\leq k$}.
\end{array}
\right.
\label{eqn-3}
\end{eqnarray}

Denote two integer sets ${\cal I}=\{i_1, i_2, \cdots$, $ i_k\}$ and $\overline{\cal I}=\{1,2, \cdots, n\}\setminus{\cal I}$. Let $A_{x_i}$ be the measurement of the party ${\cal A}_{i}, j=1, 2, \cdots, n$. Given measurements of all parties ${\cal A}_{i}$ with $i\in \overline{\cal I}$, define one quantity $I_{n,k}$ of multipartite correlations for the network shown in Fig.1 as
\begin{eqnarray}
I_{n,k}&=&\frac{1}{2^k}\sum_{x_{i},  i\in {\cal I}}\langle A_{x_{1}}A_{x_{2}}\cdots A_{x_{n}}\rangle,
\label{eqn-4}
\end{eqnarray}
where $\langle A_{x_{1}}A_{x_{2}}\cdots A_{x_{n}} \rangle=\sum_{\bf a}(-1)^{\sum_{i=1}^na_{i}}P({\bf a}|{\bf x})$ and $P({\bf a}|{\bf x})$ are defined in Eq.(\ref{eqn-2}). Similarly, using the other measurements of all parties ${\cal A}_{i}$ with $i\in \overline{\cal I}$, define the other quantity $J_{n,k}$ of multipartite correlations as
\begin{eqnarray}
J_{n,k}&=&\frac{1}{2^k}\sum_{x_{i}, i\in {\cal I}}(-1)^{\sum_{j\in {\cal I}}x_{j}}
\langle A_{x_{1}}A_{x_{2}}\cdots A_{x_{n}}\rangle.
\label{eqn-5}
\end{eqnarray}
One of the main results is that the following nonlinear inequality holds \cite{SM}:
\begin{eqnarray}
|I_{n,k}|^{\frac{1}{k}}+|J_{n,k}|^{\frac{1}{k}}\leq 1,
\label{eqn-6}
\end{eqnarray}
when a network satisfies the $k$-independence or the equivalent $k$-locality.

For quantum network of Fig.1, assume that for the observer ${\cal A}_i$ there are two-valued positive-operator-valued-measurements (POVMs) defined by Hermitian positive semidefinite operators $\{\textbf{A}^{a_i}_{x_i}, x_i=0,1\}$ with $a_i\in\{0, 1\}$, where $\textbf{A}_{x_i}$ satisfy $\sum_{x_i}\textbf{A}^{a_i}_{x_i}=\textbf{I}_i$ for each $a_i$ and $\textbf{I}_i$ is the identity operator on ${\cal A}_i$'s system, $i=1, 2, \cdots, n$. The expectation of quantum mechanical correlations among space-like separated observers are given by $\langle \otimes_{i=1}^n\textbf{A}_{x_i}\rangle=\textrm{Tr}(\otimes_{i=1}^n\textbf{A}_{x_i}\rho)$, where $\textbf{A}_{x_i}=\textbf{A}^{a_i=0}_{x_i}-\textbf{A}^{a_i=1}_{x_i}$ and $\rho$ denotes the quantum resources used in Fig.1. The second result is the following Cirel'son bound \cite{SM,Cir} (also written Tsirelson bound \cite{Tsi})
\begin{eqnarray}
|I^{q}_{n,k}|^{\frac{1}{k}}+|J^{q}_{n,k}|^{\frac{1}{k}}\leq \sqrt{2},
\label{eqn-7}
\end{eqnarray}
when quantum network has $k$ observers that do not share quantum resources, where $I^{q}_{n,k}$ and $J^{q}_{n,k}$ are the corresponding quantities of $I_{n,k}$ and $J_{n,k}$ derived from quantum mechanical correlations.

$I_{n,k}$ and $J_{n,k}$ are important quantities for characterizing a network. Given a network several inequalities can be constructed from Eq.(\ref{eqn-6}) using different $I_{n,k}$ and $J_{n,k}$, which are followed from different subsets of hidden states satisfying Eq.(\ref{eqn-3}). $k$ is another important quantity for featuring networks. When $k=1$ inequality (\ref{eqn-6}) reduces to linear Bell inequality \cite{CHSH}. Generally, a larger $k$ implies more multipartite correlations being involved in inequality (\ref{eqn-6}). So, it is reasonable to find the maximum $k$ (i.e., $k_{\max}$) and the corresponding independent parties. Unfortunately, $k_{\max}$ depends on the network configurations. Intuitively, it requires to check the independence of all subsets (exponential number) of $n$ parties ${\cal A}_{i}$s. Hence, it may be hard to get $k_{\max}$ of a general network, see Appendix B \cite{SM} for two explanations of this problem. In spite of that, analytical methods exist for some complex networks (Fig.3) beyond chain-shaped networks or star-shaped networks \cite{BGP,BRGP,TSCA,BBBC,ACSC,GMTR}. Additionally, from a suboptimal $k\leq k_{\max}$ we can construct useful inequality (\ref{eqn-6}) if $k\geq 2$. Notably, the suboptimal $k$ is equivalent to the maximal matching of the unweighted bipartite graph \cite{SM}, which can be solved using a polynomial algorithm \cite{HK,YG,ACGK}. Therefore,  inequalities (\ref{eqn-6}) can be efficiently constructed for any networks with multiple independent parties \cite{SM}.

The quantum bound in inequality (\ref{eqn-7}) is different from that in inequality (\ref{eqn-6}) for classical network in terms of the GLCM. Unfortunately, it is difficult to verify for general quantum networks. The following applications are to partially address this problem.

\begin{figure}
\begin{center}
\resizebox{95pt}{70pt}{\includegraphics{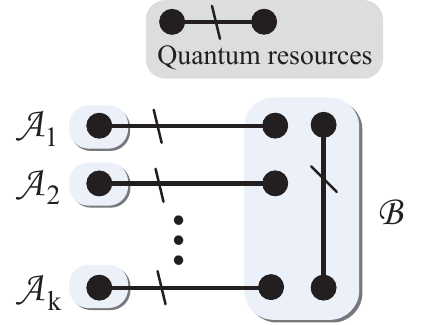}}
\end{center}
\caption{\small (Color online) Schematic quantum network with $k$ independent observers. Each pair of observers ${\cal A}_i$ and ${\cal B}$ share some entangled states, $i=1, 2, \cdots, k$. ${\cal B}$ may contain states that are not shared with ${\cal A}_i$s. ${\cal A}_i$ and ${\cal A}_j$ do not share entangled states for any $i\not=j, i, j=1, 2, \cdots, k$. The graph of two black dots connected by one long line and one short line schematically represents quantum resources.}
\label{fig-2}
\end{figure}

\textit{Generic non-multilocality of quantum networks with multiple independent observers}. The prediction of the quantum theory is incompatible with the local realism model \cite{Bel1}. This feature is generic for entangled two spin-$\frac{1}{2}$ particles \cite{CFS,Gis} or multipartite entangled states \cite{PR} using CHSH inequality \cite{CHSH}. A natural question is whether the inconsistence is typical for quantum networks. We aim to answer this question for those networks consisting of bipartite entangled pure states \cite{EPR} and generalized GHZ states \cite{GHZ} using the presented inequality (\ref{eqn-6}). Let $\textsf{G}_q=(\textsf{V, P, E})$ be a finite-size quantum network shown in Fig.1, where $\textsf{V}$ denotes all observers (nodes), $\textsf{P}$ denotes all particles of quantum resources, and $\textsf{E}$ denotes all edges (two particles are connected by one edge if they are entangled). Assume that $\textsf{G}_q$ is $k$-independent, where ${\cal A}_1, {\cal A}_2, \cdots, {\cal A}_k$ denote independent observers. There is an equivalent network shown in Fig.2, where ${\cal B}$ denotes all observers in $\textsf{G}_q$ except for ${\cal A}_i$s.  For each equivalent network, we prove the following theorem:

\textbf{Theorem A}: For any $k$-independent ($k\geq 2$) quantum network $\textsf{G}_q$, assume that the quantum resources consist of bipartite entangled pure states and generalized GHZ states. Then the following results hold:
\begin{itemize}
\item[(1)] A set of observables exists for all observers such that the multipartite quantum correlations are inconsistent with generalized local realism;
\item[(2)] A set of observables exists for all observers such that the violation of inequality (\ref{eqn-6}) is maximal when  quantum resources consist of EPR states and GHZ states.
\end{itemize}

Different from previous Bell inequalities for the chain-shaped or star-shaped network consisting of EPR states \cite{BGP,BRGP,ACSC,GMTR}, Theorem A shows that the inequalities presented in Eq.(\ref{eqn-6}) are useful for acyclic or cyclic networks consisting of bipartite entangled pure states and generalized GHZ states. Furthermore, assume that the quantum resources consist of Werner states: $\rho_w=\otimes_{i=1}^{m_1}\otimes_{j=1}^{m_2}[v_i|\Phi_i\rangle\langle \Phi_i|+(1/4-v_i/4)\mathbbm{1}_4]\otimes [w_j|\Psi_j\rangle\langle \Psi_j|+(1/2^{s_j}-w_j/2^{s_j})\mathbbm{1}_{2^{s_j}}]$, where $|\Phi_i\rangle=a_i|00\rangle+b_i|11\rangle$ are generalized EPR states,   $|\Psi_j\rangle=\hat{a}_j|0\rangle^{\otimes s_j}+\hat{b}_j|1\rangle^{\otimes s_j}$ are generalized GHZ states with $s_j\geq 3$, $m_1$ and $m_2$ denote the numbers of the respective generalized EPR states and GHZ states, $\mathbbm{1}_{2^{s_j}}$ is $2^{s_j}$ square identity matrix, and $0\leq v_i, w_j\leq 1$. We evaluate the critical viabilities as follows:

\textbf{Theorem B}: Assume that a $k$-independent ($k\geq 2$) quantum network $\textsf{G}_q$ consists of Werner states $\rho_w$, then the product of the critical visibilities $v_j^*, w^*_\jmath$ is given by
\begin{eqnarray}
\prod_{i=1}^{m_1}\prod_{j=1}^{m_2}v^*_iw^*_j
\leq \frac{1}{(1+\prod_{i=1}^{m_1}\prod_{j=1}^{m_2}(4a_ib_i\hat{a}_j\hat{b}_j)
^{\frac{2}{k}})^{\frac{k}{2}}}
\end{eqnarray}
for which the multipartite quantum correlations violate inequality (\ref{eqn-6}).

Theorems A and B hold for each $k$-independent network in Fig.2 with $2\leq k\leq k_{\max}$. Thus, several violations exist for the same quantum network with different equivalent networks. These violations provide restrictions for different multipartite quantum correlations involved in $I_{n,k}$ and $J_{n,k}$ and are valuable for further explorations.

The proof of Theorem A is to construct proper observables for all observers \cite{Gis,PR}. In fact, the observables are dependent on specific parameters \cite{SM}. In addition, all observables of the network shown in Fig.2 will be equivalently defined for all observers of the original network $\textsf{G}_q$. In particular, with these observables the maximal violations with respect to Tsirelson's bound presented in Eq.(\ref{eqn-7}) exist for EPR states and GHZ states as the quantum resources \cite{SM}. For unknown EPR states and GHZ states, our proof enables probabilistically verifying the violations \cite{SM}. Similar proof can be completed for Theorem B \cite{SM}.

\begin{figure}
\begin{center}
\resizebox{240pt}{140pt}{\includegraphics{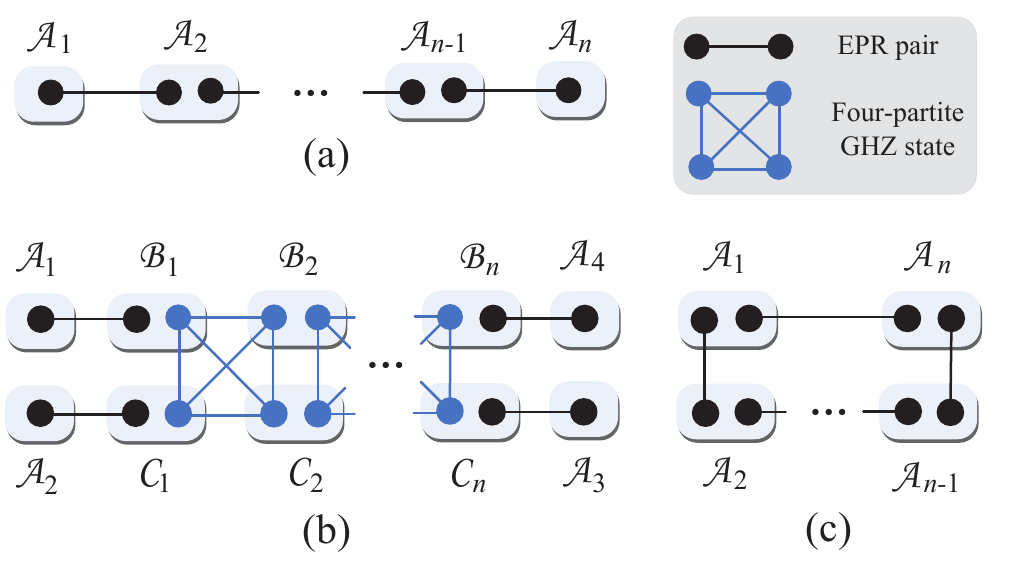}}
\end{center}
\caption{\small (Color online) (a) Long-distance entanglement distribution. All parties ${\cal A}_2, \cdots, {\cal A}_{n-1}$ jointly distribute a bipartite entangled state to two parties ${\cal A}_1$ and ${\cal A}_n$ with bipartite entangled states as quantum resources. (b) Hybrid star-shaped network. ${\cal B}_i$s and ${\cal C}_{i}$s jointly distribute a four-partite GHZ state to all parties ${\cal A}_j$s with bipartite entangled states and generalized four-partite GHZ states. (c) Cyclic network consisting of bipartite entangled states. One edge connects two entangled particles (black dots). One cubic denotes an observer who have several particles.}
\label{fig-3}
\end{figure}

\section*{Examples}

\textit{Chain-shaped network.}-Tripartite quantum correlations derived from the entanglement swapping violate inequality (\ref{eqn-6}) with $n=3$ and $k=2$ \cite{BGP,BRGP,GMTR}. The long-distance entanglement distributing generates a chain-shaped network shown in Fig.3(a). From Theorem A multipartite quantum correlations violate inequality (\ref{eqn-6}) for bipartite entangled pure states, where $k=\lceil n/2\rceil$ denotes the number of independent observers, and $\lceil x\rceil$ denotes the smallest integer no less than $x$. The maximal violation achieves for EPR states. These results answer the conjecture of verifying long-distance entanglement distributing \cite{BRGP}. Moreover, Theorem A implies similar violations for the case of multiple bipartite entangled pure states being shared by two adjoining observers. Thus our result goes beyond iterative method \cite{RBBP} that involves complicated computations for large $n$. For Werner states, the product of the critical visibilities is no less than the product of the visibility of each generalized EPR state \cite{SM}.

\textit{Hybrid star-shaped network.}-Different from previous star-shaped network \cite{BGP,BRGP,GMTR}, new network in Fig.3(b) consists of bipartite entangled states and generalized four-partite GHZ states. Theorem A implies that multipartite quantum correlations violate inequality (\ref{eqn-6}) with $k=4+\lfloor (n-1)/2\rfloor$, where $\lfloor x\rfloor$ denotes the maximal integer no more than $x$. Similar result holds for Werner states as quantum resources from Theorem B. Notably, Scarani and Gisin \cite{SG} showed some generalized GHZ states do not violate special Bell inequalities \cite{Merm,Ard,BK}. Nevertheless, all generalized GHZ states of even $n$ particles violate another Bell inequality \cite{ZBL}. Our example shows all generalized four-partite GHZ states violate inequality (\ref{eqn-6}). Generally, from Theorem A the generic violations of the multilocality hold for the networks consisting of generalized multipartite GHZ states.

\textit{Cyclic network.}-Consider a cyclic network in Fig.3(c) consisting of bipartite entangled pure states. Theorem A shows that multipartite quantum correlations violate inequality (\ref{eqn-6}) with $k=\lfloor\frac{n}{2}\rfloor$ when $n\geq4$. It is also maximal with respect to Tsirelson's bound given in Eq.(\ref{eqn-7}) for EPR states. Similar result holds for Werner states from Theorem B. This is the first example of nontrivial cyclic network discussed so far. We further present some examples such as butterfly network or boat network containing two or more cyclic subnetworks \cite{SM}, which are interesting in communications \cite{ACLY,Hay}.

\section*{Discussion}

For testing the non-multilocality of quantum networks consisting of general noisy resources, we provide one sufficient condition that all coefficients of $\sigma_x\otimes \cdots \otimes\sigma_x$ and $\sigma_z\otimes \cdots \otimes\sigma_z$ in quantum states are no smaller than $\frac{1}{\sqrt{2}}$, see Appendix G \cite{SM}, where $\sigma_x, \sigma_z$ are Pauli matrices. Further investigations are valuable for the non-multilocality and entanglement witness \cite{HHH}. When multiple outputs and inputs are required for observers, the linear expansion of dichotomic inputs and outputs \cite{BRGP} are inefficient to characterize all multipartite quantum correlations \cite{CG}. The general representations of $I_{n,k}$ and $J_{n,k}$ are related to the famous conjecture of the Hadamard matrix \cite{AK}. This raises two interesting problems: (1) how to characterize networks consisting of high-dimensional quantum states; (2) how to characterize cyclic quantum networks without $k\geq 2$ independent observers \cite{BRGP,TSCA,BBBC,ACSC,LS,Cha,RBBP,GMTR}, where several nontrivial examples are triangle network, symmetric cyclic network and door-type network consisting of EPR states and multipartite GHZ states, see Appendix H \cite{SM}, which maybe interesting for quantum nonlocal games \cite{SMA,Bus}.

In addition to interesting applications such as the randomness amplification, interactive proofs and quantum games \cite{Ek,ABGM,PAM,BCMD,BGK}, quantum networks allow multipartite tasks. One notable problem is to address the supremacy of quantum networks in the case of multipartite interactive proofs or computational complexities. Its improvement may provide further relevance of quantum networks and classical problems \cite{BGK}.

In conclusion, we presented explicit nonlinear Bell-inequalities for networks with multiple sources. These inequalities are computationally efficient and are used to prove the generic non-multilocality of quantum networks with independent observers. The result holds for any bipartite entangled pure states and generalized GHZ states as quantum resources. The violations are maximal with respect to Tsirelson's bound for EPR states and GHZ states. Furthermore, the upper bounds of the critical visibilities are presented for Werner states. Our results may stimulate investigators to employ the non-multilocality for quantum information processing or quantum Internet.

\section*{Acknowledgements}

We thank the helpful discussions of Luming Duan, Yaoyun Shi, Huiming Li, Xiubo Chen, Yixian Yang, Xiaojun Wang, Yuan Su. This work was supported by the National Natural Science Foundation of China (Nos.61772437,61702427), Sichuan Youth Science and Technique Foundation (No.2017JQ0048), Fundamental Research Funds for the Central Universities (No.2682014CX095), Chuying Fellowship, and CSC Scholarship.

\section*{Appendix A1:  Proof of Bell inequality (6)}

In this section, we prove Bell inequality (6) for a general network in terms of the generalized locally causal model. From the definition of $k$-locality given in Eq.(3), $P(a_1, a_2, \cdots, a_n|x_1, x_2, \cdots, x_n)$ has the decomposition shown in Eq.(2). Let $\langle A_{x_{1}}A_{x_{2}}\cdots A_{x_{n}} \rangle=\sum_{a_1, a_2, \cdots, a_n}(-1)^{\sum_{i=1}^na_{i}}P(a_1, a_2, \cdots, a_n|x_1, x_2, \cdots, x_n)$. Define the expectation of the outcomes of $A_{x_i}$ as
\begin{align*}
\langle A_{x_i}\rangle=\sum_{a_i=0}^1(-1)^{a_i}P(a_i|x_i,\Lambda_i),
\tag{A1}
\end{align*}
where $i=1, 2, \cdots, n$.

Denote the integer sets ${\cal I}=\{i_1, i_2, \cdots$, $ i_k\}$ and $\overline{\cal I}=\{1,2, \cdots, n\}$ $/\{i_1, i_2, \cdots, i_k\}$. Using the inequalities $|\langle A_{x_i}\rangle|\leq 1$ for $i=1, 2, \cdots, n$, from Eqs.(4), (5) and (A1) we obtain that
\begin{align*}
|I_{n,k}|=&\frac{1}{2^k}\int_\Omega d\mu_1(\Lambda_{i_1})d\mu_2(\Lambda_{i_2})\cdots d\mu_k(\Lambda_{i_k})
\prod_{i_s\in {\cal I}}|\langle A_{x_{i_s}=0}\rangle+\langle A_{x_{i_s}=1}\rangle| \prod_{j\in \overline{\cal I}}|\langle A_{j}\rangle|
\nonumber
\\
\leq &\frac{1}{2^k}\int_\Omega d\mu_1(\Lambda_{i_1})d\mu_2(\Lambda_{i_2})\cdots d\mu_k(\Lambda_{i_k})
\prod_{i_s\in {\cal I}}|\langle A_{x_{i_s}=0}\rangle+\langle A_{x_{i_s}=1}\rangle|.
\tag{A2}
\end{align*}

By setting $\langle\Delta^\pm A_{x_{i_s}}\rangle=\frac{1}{2}(\langle A_{x_{i_s}=0}\rangle\pm \langle A_{x_{i_s}=1}\rangle)$, Eq.(A2) yields to
\begin{align*}
|I_{n,k}|\leq &\int_\Omega d\mu_1(\Lambda_{i_1})d\mu_2(\Lambda_{i_2})\cdots d\mu_k(\Lambda_{i_k})
 \prod_{i_s\in {\cal I}}|\langle \Delta^+ A_{x_{i_s}}\rangle|
\nonumber
\\
\leq &
\prod_{i_s\in {\cal I}}\int_{\Omega'_s}d\mu(\Lambda_{i_s})|\langle \Delta^+ A_{x_{i_s}}\rangle|,
\tag{A3}
\end{align*}
where $\mu(\Lambda_{i_s})=\prod_{\lambda_j\in \Lambda_{i_s}}\mu_j(\lambda_j)$ and $\Omega_s'=\times_{\lambda_j\in \Lambda_{i_s}}\Omega_j$ (the product space of probability spaces  $\Omega_j$s), $s=1, 2, \cdots, k$.

Similarly, we obtain that
\begin{align*}
|J_{n,k}|=&|\int_\Omega d\mu_1(\Lambda_{i_1})d\mu_2(\Lambda_{i_2})\cdots d\mu_k(\Lambda_{i_k})
\prod_{i_s\in {\cal I}}\langle \Delta^- A_{x_{i_s}}\rangle\prod_{j\in \overline{\cal I}}\langle A_{j}\rangle|
\\
\leq & \prod_{i_s\in {\cal I}}\int_{\Omega_s'} d\mu(\Lambda_{i_s})|\langle \Delta^- A_{x_{i_s}}\rangle|.
\tag{A4}
\end{align*}

Using the Mahler inequality \cite{Mah}, from the inequalities (A3) and (A4) we get that
\begin{align*}
 |I_{n,k}|^{\frac{1}{k}}+|J_{n,k}|^{\frac{1}{k}}
\leq &(\prod_{i_s\in {\cal I}}\int_{\Omega_s'}d\mu(\Lambda_{i_s})
(|\langle \Delta^+ A_{x_{i_s}}\rangle|
 +|\langle \Delta^- A_{x_{i_s}}\rangle|))^{\frac{1}{k}}
\nonumber\\
\leq &(\prod_{i_s\in {\cal I}}\int_{\Omega_s'}d\mu(\Lambda_{i_s}) )^{\frac{1}{k}}
\tag{A5}\\
=&1,
\tag{A6}
\end{align*}
where the inequality (A5) is from the inequalities $|\langle \Delta^+ A_{x_{i_s}}\rangle|+|\langle \Delta^- A_{x_{i_s}}\rangle|=\max\{|\langle A_{x_{i_s}=0}\rangle|$, $|\langle A_{x_{i_s}=1}\rangle|\}\leq 1$ for $s=1, 2, \cdots, n$; and Eq.(A6) is from the normalization condition of the probability distribution of hidden states.

\section*{Appendix A2: Proof of the inequality (7)}

In this subsection, we prove the Tsirelson's bound presented in Eq.(7). For a network shown in Fig.1, assume that there are two-valued positive-operator-valued-measurements (POVMs) ${\bf A}_{x_1}, {\bf A}_{x_2}, \cdots, {\bf A}_{x_n}$ with $x_i\in \{0,1\}$, where ${\bf A}_{x_i=0}$ and ${\bf A}_{x_i=1}$ are defined on the subsystem of the $i$-th observer and the outcomes of them are labeled by $\pm1$. Here, a POVM can be probabilistically realized by performing the projective measurements on a larger quantum system according to the Neumark dilation theorem \cite{Paul}. Note that these operators satisfy the commutativity condition $[{\bf A}_{x_i},{\bf A}_{x_j}]=0$ for $i\not=j$ because they are performed on different subsystems. The expectation of quantum mechanical correlations among space-like separated observers are given by $\langle \otimes_{i=1}^n\textbf{A}_{x_i}\rangle={\rm Tr}(\otimes_{i=1}^n\textbf{A}_{x_i}\rho)$, where $\rho$ denotes the joint system of quantum resources used in Fig.1.

We firstly prove the following lemma (which may be mathematically presented in some papers because of its simplicity)

{\bf Lemma 1}. For any $\theta_1, \theta_2, \cdots, \theta_n\in [0, \pi]$ and  integer $n\geq 2$, we obtain that the following inequality
 \begin{align*}
(\prod_{i=1}^n\sin\theta_i)^{\frac{1}{n}}\leq \sin(\frac{1}{n}\sum_{i=1}^n\theta_i),
\tag{A7}
\end{align*}
where the equality holds if and only if $\theta_1=\theta_2=\cdots =\theta_n$.

{\bf Proof}. The proof is completed by induction. For $n=2$, the inequality (A7) is equivalent to
\begin{align*}
\sin\theta_1\sin\theta_2\leq &\sin^2(\frac{\theta_1+\theta_2}{2})
\\
=&\frac{1}{2}(1-\cos(\theta_1+\theta_2))
\\
=&\frac{1}{2}(1-\cos\theta_1\cos\theta_2+\sin\theta_1\sin\theta_2)
\tag{A8}
\end{align*}
which implies that $\cos(\theta_1-\theta_2)\leq 1$. This is satisfied for   any $\theta_1, \theta_2\in [0, \pi]$.

Now, assume that for any $n$ with $n\leq k-1$, the inequality (A7) holds all  $\theta_1, \theta_2, \cdots, \theta_n\in [0, \pi]$. For even $n=k$, from the assumption we obtain that
\begin{align*}
(\prod_{i=1}^{n}\sin\theta_i)^{\frac{1}{n}}
= &\sqrt{(\prod_{i=1}^{m}\sin\theta_i)^{\frac{1}{m}}}
\sqrt{(\prod_{i=m}^{n}\sin\theta_i)^{\frac{1}{m}}}
\\
\leq &\sqrt{\sin(\frac{1}{m}\sum_{i=1}^{m}\theta_i)}
\sqrt{\sin(\frac{1}{m}\sum_{i=m+1}^{n}\theta_i)}
\tag{A9}
\\
\leq &\sin(\frac{1}{n}\sum_{i=1}^{n}\theta_i),
\tag{A10}
\end{align*}
where the equality in Eq.(A9) holds if and only if
$\theta_1=\theta_2=\cdots=\theta_m$ and $\theta_{m+1}=\theta_{m+2}=\cdots=\theta_n$, and the equality in Eq.(A10) holds if and only if $\sum_{i=1}^m\theta_i=\sum_{j=m+1}^{n}\theta_j$, and $m=\frac{n}{2}$. So, the equality in Eq.(A7) holds if and only if $\theta_1=\theta_2=\cdots=\theta_n$.

For odd $n$ with $n=k$, by introducing an ancillary variable $\theta_{n+1}$, from the assumption we obtain that
\begin{align*}
(\prod_{i=1}^{n+1}\sin\theta_i)^{\frac{1}{n+1}}
= &\sqrt{(\prod_{i=1}^{m}\sin\theta_i)^{\frac{1}{m}}}
\sqrt{(\prod_{i=m+1}^{n+1}\sin\theta_i)^{\frac{1}{m}}}
\\
\leq &\sqrt{\sin(\frac{1}{m}\sum_{i=1}^{m}\theta_i)}\sqrt{\sin(\frac{1}{m}
\sum_{i=m+1}^{n+1}\theta_i)}
\tag{A11}
\\
\leq &\sin(\frac{1}{n+1}\sum_{i=1}^{n+1}\theta_i),
\tag{A12}
\end{align*}
where the equality in Eq.(A11) holds if and only if
$\theta_1=\theta_2=\cdots=\theta_m$ and $\theta_{m+1}=\theta_{m+2}=\cdots=\theta_n$, and the equality in Eq.(A12) holds if and only if $\sum_{i=1}^m\theta_i=\sum_{j=m+1}^{n+1}\theta_j$, and $m=\frac{n+1}{2}$. Now, by setting $\theta_{n+1}=\frac{1}{n}\sum_{i=1}^n\theta_i$, we get $\theta_{n+1}=\frac{1}{n+1}\sum_{i=1}^{n+1}\theta_i$. Thus, the inequality (A12) yields that
\begin{align*}
(\prod_{i=1}^{n}\sin\theta_i)^{\frac{1}{n+1}}
(\sin\theta_{n+1})^{\frac{1}{n+1}}\leq
&\sin\theta_{n+1},
\end{align*}
which implies the inequality (A7). The equality in Eq.(A7) holds if and only if $\theta_1=\theta_2=\cdots=\theta_n$. $\square$

Now, we continue to prove the inequality (7). For the sake of simplicity, let ${\cal I}=\{1, 2, \cdots, k\}$. Denote $\|{\bf X}\|={\rm sup}\{{\rm Tr}({\bf X\rho})$, $\rho\in \mathbb{H} \mbox{ with } {\rm Tr} \rho=1\}$ as the norm of a positive semidefinite operator ${\bf X}$ on Hilbert space $\mathbb{H}$. From Eqs.(4) and (5), the inequalities $\|{\bf A}_{x_j}\|\leq 1$, and the linearity of the expectation operation $\langle \cdot \rangle$, we obtain that
\begin{align*}
F:=&|I^q_{n,k}|^{\frac{1}{k}}+|J^q_{n,k}|^{\frac{1}{k}}
\\
=&|\sum_{x_1, x_2, \cdots, x_k=0,1}
\frac{1}{2^k}\langle (\otimes_{j=1}^k{\bf A}_{x_j})\otimes (\otimes_{s\in \overline{\cal I}}{\bf A}_{x_s})\rangle|^{\frac{1}{k}}
\\
&+|\sum_{x_1, x_2, \cdots, x_k=0,1}(-1)^{\sum_{j=1}^kx_j}
\frac{1}{2^k}\langle (\otimes_{j=1}^k{\bf A}_{x_j})\otimes (\otimes_{s\in \overline{\cal I}}{\bf A}_{x'_s})\rangle|^{\frac{1}{k}}
\\
\le&\frac{1}{2}|\langle  \otimes_{j=1}^k({\bf A}_{j,0}+{\bf A}_{j,1})\rangle|^{\frac{1}{k}}
+\frac{1}{2}|\langle \otimes_{j=1}^k({\bf A}_{j,0}-{\bf A}_{j,1})\rangle|^{\frac{1}{k}},
\tag{A13}
\end{align*}
where ${\bf A}_{j,0}={\bf A}_{x_j=0}$, ${\bf A}_{j,1}={\bf A}_{x_j=1}$, and $x_s, x'_s \in \{0,1\}$, $j=1, 2, \cdots, n$.

Moreover, using the commutativity conditions $[{\bf A}_{x_i}, {\bf A}_{x_j}]=0$ for any $i\not=j$, the inequality (A13) yields to
\begin{align*}
F^2\leq &\frac{1}{4}|\langle
 \otimes_{j=1}^k({\bf A}_{j,0}+{\bf A}_{j,1})\rangle|^{\frac{2}{k}}
+\frac{1}{4}|\langle \otimes_{j=1}^k({\bf A}_{j,0}-{\bf A}_{j,1})\rangle|^{\frac{2}{k}}
 +\frac{1}{2}|\langle \otimes_{j=1}^k [{\bf A}_{j,0},{\bf A}_{j,1}]\rangle|^{\frac{1}{k}}
\\
\leq &
\frac{1}{4}|\langle \otimes_{j=1}^k(2{\bf I}+{\bf B}_j)\rangle|^{\frac{1}{k}}
+\frac{1}{4}|\langle \otimes_{j=1}^k(2{\bf I}-{\bf B}_j)\rangle|^{\frac{1}{k}}
+1
\tag{A14}
\end{align*}
from the operator inequalities ${\bf A}_{j,0}^2\leq {\bf I}, \textbf{A}_{j,1}^2\leq {\bf I}$ and the inequalities $\|[{\bf A}_{j,0},{\bf A}_{j,1}]\|\leq 2\|{\bf A}_{j,0}\|\cdot\| {\bf A}_{j,1}\|\leq 2$, where ${\bf B}_j={\bf A}_{j,0}{\bf A}_{j,1}+{\bf A}_{j,1}{\bf A}_{j,0}$, $[{\bf A}_{j,0},{\bf A}_{j,1}]={\bf A}_{j,0}{\bf A}_{j,1}-{\bf A}_{j,1}{\bf A}_{j,0}$, and ${\bf I}$ denotes the identity operator.

Note that $2{\bf I}\pm {\bf B}_j\geq 0$ (operator inequalities) because of $\|{\bf B}_j\|\leq 2$. The inequality (A14) is equivalent to
\begin{align*}
F^2\leq &
\frac{1}{4}\prod_{j=1}^k\langle 2{\bf I}+\textbf{B}_j\rangle^{\frac{1}{k}}
+\frac{1}{4}\prod_{j=1}^k\langle 2{\bf I}-\textbf{B}_j\rangle^{\frac{1}{k}}
+1
\\
= &
\frac{1}{4}\prod_{j=1}^k(2+\langle\textbf{B}_j\rangle)^{\frac{1}{k}}
+\frac{1}{4}\prod_{j=1}^k(2-\langle\textbf{B}_j\rangle)^{\frac{1}{k}}
+1.
\tag{A15}
\end{align*}

By setting $2+\langle\textbf{B}_j\rangle=4\sin^2\theta_j$ with $\theta_j\in [0,\pi]$, we obtain that $2-\langle\textbf{B}_j\rangle=4\cos^2\theta_j$, $j=1, 2, \cdots, n$. From the inequality (A15) we get that
\begin{align*}
F^2\leq &
(\prod_{j=1}^k\sin\theta_j)^{\frac{2}{k}}
+(\prod_{j=1}^k\sin(\frac{\pi}{2}-\theta_j))^{\frac{2}{k}}
+1
\\
\leq&
\sin^{2}(\sum_{j=1}^k\theta_j)
+\sin^{2}(\frac{\pi}{2}-\sum_{j=1}^k\theta_j)
+1
\tag{A16}
\\
=&2,
\tag{A17}
\end{align*}
where the inequality (A16) is from the presented Lemma above, and the equality in Eq.(A16) is from the equalities $|\sin(\frac{\pi}{2}-\sum_{j=1}^k\theta_j)|=|
\cos(\sum_{j=1}^k\theta_j)|$ and $\sin^2\theta+\cos^2\theta=1$. So, $F\leq \sqrt{2}$.

\section*{Appendix B: The number $k$ of independent parties in networks}

\subsection*{Appendix B1: The maximum $k_{\max}$}

In this subsection, we show the hardness of finding the maximum $k_{\max}$ for a general network.

The following procedure starts from an equivalent bipartite graph (in which all the parties have not been decomposed) $\textsf{G}=(\textsf{S, A, E})$ of a given network in Fig.1. $\textsf{S}$ denotes the set of $m$ independent sources $S_1, S_2, \cdots, S_m$. $\textsf{R}$ denotes the set of $n$ parties ${\cal A}_1, {\cal A}_2, \cdots, {\cal A}_n$. $\textsf{E}$ denotes the set of all edges that schematically represent the relationships of sources and parties, i.e., the edge $S_iR_j$ schematically represent the fact that the party $R_j$ receives one hidden state from the source $S_i$. Denote $\ell_i$ as the number of independent sources that are connected to the party ${\cal A}_i$, $i=1, 2, \cdots, n$. The problem of finding the maximum $k_{\max}$ can be mathematically formulated as an integer optimization as follows:
\begin{itemize}
\item[] {\bf Maximize}: $\sum_{i=1}^n y_i$

\item[] {\bf subject to}:
\begin{itemize}
\item[] $\sum_{i=1}^nx_{ij}=1$, $j=1, 2, \cdots, m$;
\item[] $\sum_{j=1}^mx_{ij}=\ell_i y_i$, $i=1, 2, \cdots, n$;
\item[] $y_i$, $x_{ij}\in \{0, 1\}$, $1\leq \ell_i\leq m$, $i=1, 2, \cdots, n$; $j=1, 2, \cdots, m$.
\end{itemize}
\end{itemize}
Here, $y_i$ is the characteristic function of the party ${\cal A}_i$, i.e., $y_i=1$ if the party ${\cal A}_i$ is included in independent parties for evaluating $k_{\max}$; Otherwise $y_i=0$. The first condition is used to ensure that each source distributes a hidden state to one party. The second condition is used to ensure that the party ${\cal A}_i$ is included in  independent parties, i.e., the number of nonzero $x_{ij}$ should equal to that of the edges connected to ${\cal A}_i$, $j=1, 2, \cdots, m$. Note that the integer optimization problem is generally NP-hard \cite{Kar}. So, we believe that the problem of evaluating $k_{\max}$ is also hard. Of course, there exists P-hard subsets of integer optimizations for special networks (see Fig.3).

In fact, in the next subsection we show that finding the maximum $k_{\max}$ of a general network is related to finding the maximum matching of a bipartite graph, which is known as a NP-hard problem \cite{West}.

\begin{figure}
\begin{center}
\resizebox{125pt}{80pt}{\includegraphics{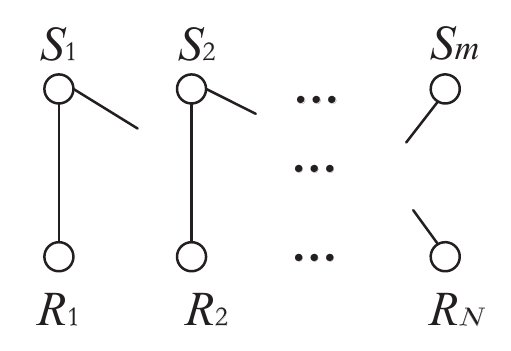}}
\end{center}
\caption{\small (Color online) The equivalent unweighted bipartite graph. $R_1, R_2, \cdots, R_{N}$ denote all decomposed parties of ${\cal A}_1, {\cal A}_2, \cdots, {\cal A}_n$, where each party ${\cal A}_{j}$ is decomposed into $\ell_j$ different new parties who have only one connected edge, and $N=\sum_{i=1}^n\ell_i$.}
\label{fig-3}
\end{figure}

\subsection*{Appendix B2: Efficiently constructing Bell inequality presented in Eq.(6)}

\begin{figure}
\begin{center}
\resizebox{235pt}{190pt}{\includegraphics{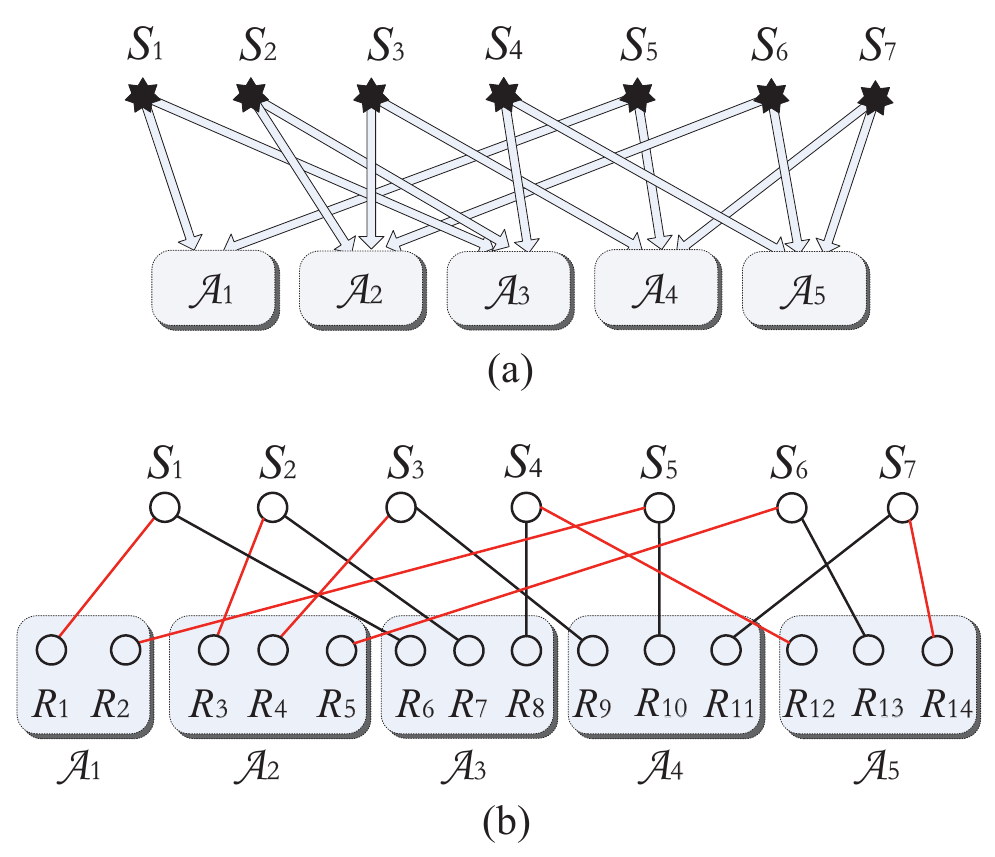}}
\end{center}
\caption{\small (Color online) (a) Simple network with 7 independent sources $S_1, S_2, \cdots, S_7$. Each star denotes one source. (b) Equivalent unweighted bipartite graph. Here, ${\cal A}_1,  {\cal A}_2, \cdots, {\cal A}_5$ are decomposed into $\{R_1, R_2\}$, $\{R_3, R_4, R_5\}$, $\{R_6, R_7, R_{8}\}$, $\{R_9, R_{10}, R_{11}\}$, $\{R_{12}, R_{13}$, $R_{14}\}$, respectively. Each circle denotes a vertex. One maximal matching of the graph consists of red edges.}
\label{fig-3}
\end{figure}

In this subsection, we present the detailed procedure to efficiently construct Bell inequality for any network shown in Fig.1. We firstly find the independent parties. And then take use of quantities $I_{n,k}$ and $J_{n,k}$ to build the desired Bell inequality (6). The key is to obtain the number $k$ of independent parties and the corresponding independent parties. Although evaluating the maximum $k_{\max}$ is a hard problem for general networks, fortunately, there exist computationally efficient algorithms to find a suboptimal $k$ or possible values of $k$. The detailed algorithm is shown in Algorithms 1 and 2.

We firstly present an example shown in Fig.S2 to explain the main idea. Assume that there are five parties ${\cal A}_1, {\cal A}_2, \cdots$, ${\cal A}_5$ who receive states from independent sources $S_1, S_2$, $\cdots$, $S_7$ shown in Fig.S2(a). The construction is divided into the following four steps:
\begin{itemize}
\item[S1] According to different sources, we decompose the party ${\cal A}_1$ into two parties $R_1$ and $R_2$, where $R_1$ receives one state from the source $S_1$ while $R_2$ receives one state from the source $S_5$. Similarly, we decompose all the other parties ${\cal A}_2, {\cal A}_3, \cdots, {\cal A}_5$ into new parties $\{R_3, R_4, R_5\}$, $\{R_6, R_7, R_{8}\}$, $\{R_9, R_{10}, R_{11}\}$, $\{R_{12}, R_{13}$, $R_{14}\}$, respectively, where each party $R_{i}$ receives only one state from some source $S_j$. And then, we obtain an equivalent unweighted bipartite graph shown in Fig.S2(b), where the upper vertexes consist of all the sources $S_1, \cdots, S_{7}$, the lower vertexes consist of all the parties $R_1, \cdots, R_{14}$, and each edge $S_iR_j$ schematically represents the fact that the party $R_j$ receives one state from the source $S_i$.
\item[S2] Using Hopcroft-Karp algorithm \cite{ddd} we find one maximal matching of the graph consisting of red edges $S_1R_1, S_5R_2, S_2R_3, S_3 R_4, S_6 R_5, S_4 R_{12}, S_7R_{14}$ shown in Fig.S2(b), where a matching is a subset of the edges satisfying that no two edges share a vertex \cite{West}.
\item[S3] Choose all the parties ${\cal A}_1, {\cal A}_2, {\cal A}_5$ who are connected at least one red edge. By checking the completeness of three parties ${\cal A}_1, {\cal A}_2, {\cal A}_5$ (all the decomposed parties of each party ${\cal A}_i$ are connected by red edges), we obtain ${\cal A}_1$, ${\cal A}_2$ as two independent parties.
\item[S4] According to Eq.(6) in the main text, we construct a nonlinear Bell inequality as
\begin{align*}
\sqrt{|I_{5,2}|}+\sqrt{|J_{5,2}|}\leq 1,
\tag{B1}
\end{align*}
where $I_{5,2}=\frac{1}{4}\sum_{x_{1}, x_2=0,1}\langle A_{x_{1}}A_{x_{2}}A_{x_3}A_{x_4}A_{x_{5}}\rangle$ and $J_{5,2}=\frac{1}{4}\sum_{x_{1}, x_2=0,1}(-1)^{x_{1}+x_2} \langle A_{x_{1}}A_{x_{2}}A_{x'_3}A_{x'_4}A_{x'_{5}}\rangle$, $x_j$ and $x_j'$ are different bits, $j=3, 4, 5$.
\end{itemize}

\begin{algorithm}[htbp]
\caption{Constructing Bell inequalities for each network $\textsf{N}$ shown in Fig.1}
\KwIn{Finite-size network $\textsf{N}$ in terms of the GLCM}
\KwOut{Nonlinear Bell inequalities (6)}
\begin{itemize}
\item[(i)] Obtain an equivalent bipartite graph $\textsf{G}=(\textsf{S, R, E})$ shown in Fig.S1.

\item[(ii)] Find a maximal matching $\textsf{E}'\subseteq \textsf{E}$ of bipartite graph $\textsf{G}=(\textsf{S, R, E})$ using Hopcroft-Karp algorithm \cite{ddd}.

\item[(iii)]Find the number $k$ of independent parties and the corresponding independent parties ${\cal A}_{i_j}$s of the network $\textsf{N}$ shown in Fig.1.

\item[(iv)]Construct Bell inequalities (6) by taking use of the quantities $I_{n,k}$ and $J_{n,k}$ defined in Eqs.(4) and (5) respectively if $k\geq 2$; Otherwise, repeating (i)-(iii).
\end{itemize}
\end{algorithm}

To complete the step (i) of Algorithm 1, we schematically decompose each party ${\cal A}_{i}$ into $\ell_i$ different parties $R_{t_{i-1}+1}, R_{t_{i-1}+2} \cdots, R_{t_{i}}$, where each party $R_{t_{i-1}+j}$ receives only one state from some source, and $\ell_i$ denotes the number of independent sources that distribute states to the party ${\cal A}_i$, $N=\sum_{j=1}^n\ell_j$. And then, we regard all independent sources $S_1, S_2, \cdots, S_m$ of the network shown in Fig.1 as upper vertices in the set $\textsf{S}$ while all decomposed parties $R_{1}, R_2, \cdots, R_{N}$ are regarded as lower vertices in the set $\textsf{R}$. Each edge $S_iR_j$ schematically represents the fact that the source $S_i$ distributes one state to the decomposed party $R_j$.

In the step (ii) of Algorithm 1, a matching is a subset of the edges satisfying that no two edges share a vertex \cite{ddd}. For a network shown in Fig.1, there is an unweighted bipartite graph $\textsf{G}$ shown in Fig.S1 from the step (i). And then, by using Hopcroft-Karp algorithm \cite{ddd} we can find a maximal matching $\textsf{E}'$ of $\textsf{G}$.

To complete the step (iii), we firstly find all original parties ${\cal A}_i$s who have at least one decomposed party in the vertex set of the maximal matching $\textsf{E}'$. Denote $\textsf{A}$ as the desired set of all these parties ${\cal A}_i$s. And then, we check the completeness of each party ${\cal A}_i\in \textsf{A}$, where the completeness means that the vertex set of $\textsf{E}'$ contains all decomposed parties of ${\cal A}_i$. Note that the number $k$ of independent parties in Fig.1 equals to the number of original parties satisfying the completeness. Hence, for each maximal matching of the unweighted bipartite graph $\textsf{G}$, there may exist an integer $k\geq 2$ and the corresponding independent parties. Otherwise, another maximal matching should be found.

For a network shown in Fig.1, there is an unweighted bipartite graph $\textsf{G}$ shown in Fig.S1. Moreover, for each maximal matching of $\textsf{G}$, from the steps (ii) and (iii), we obtain the number $k$ of independent parties ${\cal A}_i$s, where their sources satisfy Eq.(3). Hence, when $k\geq 2$ from Eqs.(4)-(6) we obtain a nonlinear Bell inequalities. Conversely, for a network shown in Fig.1 satisfying Eq.(3),
consider one subnetwork ${\cal G}_1$ consisting of all independent parties and sources, and the other subnetwork ${\cal G}_2$ consisting of the party ${\cal B}$ and all the edges connected to it in Fig.2. For ${\cal G}_i$, from the step (i) we can easily construct the corresponding unweighted bipartite graph $\textsf{G}_i$, $i=1,2$. For the independent assumption in Eq.(3), $\textsf{G}_1$ is a fully disconnected graph that has no adjacent edges. Otherwise, there are two parties ${\cal A}_i$ and ${\cal A}_j$ who share at least one source. If $\textsf{G}_1$ contains all sources $S_i$s, then it is a maximal matching of the unweighted bipartite graph $\textsf{G}$ that is the equivalent graph of the network shown in Fig.1. Otherwise, there are some sources $S_{i_1}, S_{i_2} \cdots, S_{i_s}$ that do not distribute states for all parties ${\cal A}_1, {\cal A}_2, \cdots, {\cal A}_k$. In this case, we can find a maximal matching $\textsf{G}'_2$ of $\textsf{G}_2$ using a polynomial-time algorithm \cite{ddd}. Note that $\textsf{G}_1\cup\textsf{G}'_2$ is a maximal matching of the unweighted bipartite graph $\textsf{G}$. So, we have shown that evaluating the number $k$ of independent parties in Fig.1 is equivalent to finding a maximal matching of the equivalent unweighted bipartite graph. Specially, finding the maximum $k_{\max}$ is related to finding the maximum matching of the equivalent unweighted bipartite graph, i.e., a matching that contains the largest possible number of edges. Unfortunately, the maximum matching problem is generally NP-hard \cite{ddd}.

The steps (i)-(iii) of Algorithm 1 are used to obtain the number $k$ of independent parties and the corresponding independent parties ${\cal A}_{i_j}s$ of the network $\textsf{N}$ shown in Fig.1. Assume that the network $\textsf{N}$ has $N$ schematic links, where each link represents the relationship that one source distributes one state to a party. The time complexity of the step (i) is $O(N)$ which is from $N$ decomposition operations. The step (ii) is the most difficult part of Algorithm 1. Fortunately, we can take use of Hopcroft-Karp algorithm \cite{ddd} that is a polynomial time algorithm ($O(n^{2.5})$ for dense graphs) to get a maximal matching of $\textsf{G}$. The time complexity of the step (iii) is no more than $O(n^2)$. The time complexity of the step (iv) is trivial for a given number $k$ and the corresponding independent parties. So, the total time complexity is bounded by $O(n^{2.5})$. It means that Algorithm 1 has polynomial-time complexity if $k\geq 2$. Algorithm 1 cannot ensure us to get at least one nonlinear Bell inequality in deterministic polynomial-time because one may get $k=1$ for some maximal matchings in the worst case. However, it is efficient for networks with large numbers of independent parties.

To make up the disadvantage of Algorithm 1 for small $k$, we need the following special algorithm. For a general network $\textsf{N}$ with at least $k$ independent parties, there is another accessible method to get a small $k\geq 2$.
\begin{algorithm}[h]
\caption{Constructing Bell inequalities for each network $\textsf{N}$ shown in Fig.1}
\KwIn{Finite-size network $\textsf{N}$ in terms of the GLCM}
\KwOut{Nonlinear Bell inequalities (6)}
\begin{itemize}
\item[(i)] Randomly label all parties as ${\cal A}_1, {\cal A}_2, \cdots, {\cal A}_n\in \textsf{S}$ and all independent sources as $\lambda_1, \lambda_2, \cdots, \lambda_m\in \Lambda_0$. Denote $\Lambda_{j}$ as the set of sources related to the party ${\cal A}_{j}$, $j=1, \cdots, n$;
\item[(ii)] Choose $k$ different parties ${\cal A}_{i_1}, {\cal A}_{i_2}, \cdots, {\cal A}_{i_k}\in \textsf{S}$. Output $k$ and ${\cal A}_{i_1}, {\cal A}_{i_2}, \cdots, {\cal A}_{i_k}$ if all sets $\Lambda_{i_1}, \Lambda_{i_2}, \cdots, \Lambda_{i_k}$ satisfy the $k$-locality condition shown in Eq.(3); Otherwise, repeat this step by choosing another $k$ different parties;
\item[(iii)]Construct Bell inequalities (6) by taking use of the quantities $I_{n,k}$ and $J_{n,k}$ defined in Eqs.(4) and (5) respectively.
\end{itemize}
\end{algorithm}

Note that in Algorithm 2 the time complexity of the step (ii) is bounded by $O((^n_k))$, where $(^n_k)$ denotes the binomial coefficient that is a polynomial function of $n$ when $k$ is a small integer. Moreover, there always exist $k$ different parties ${\cal A}_{i_1}, {\cal A}_{i_2}, \cdots, {\cal A}_{i_k}\in \textsf{S}$ satisfying the $k$-locality condition shown in Eq.(3) when the network $\textsf{N}$ has at least $k$ independent parties. Thus, Algorithm 2 provides a deterministic polynomial time algorithm to construct nonlinear Bell inequalities (6) with small $k$. Hence, From Algorithms 1 and 2, we can efficiently construct nonlinear Bell inequalities (6).

\section*{Appendix C: Proof of Theorem A}

In this section, inspired by the methods in \cite{3,4} we prove Theorem A for a network shown in Fig.1 with quantum resources consisting of bipartite entangled pure states and GHZ states. In the following experiment of verifying the non-multilocality, after all parties contained in the party ${\cal B}$ perform some measurements depending on their input bits $y$ on their particles and obtain output bits $b$, all parties ${\cal A}_1, {\cal A}_2, \cdots, {\cal A}_k$ perform some measurements depending on their input bits $x_1, x_2, \cdots, x_k$ on their particles and obtain output bits $a_1, a_2, \cdots, a_k$, respectively, where $a_i, b, x_i, y\in \{0,1\}$.

The proof is completed by following the procedure from special quantum resources to general quantum resources.

\subsection*{Appendix C1: EPR states as quantum resources}

In this subsection, we assume that the quantum resources consist of generalized Einstein-Podolsky-Rosen (EPR) states \cite{EPR}:
\begin{align*}
|\Xi\rangle=\otimes_{i=1}^{m}|\Phi_i\rangle,
\tag{C1}
\end{align*}
where $|\Phi_i\rangle=a_{i}|00\rangle+b_i|11\rangle$ are generalized EPR states with real coefficients $a_i, b_i$ satisfying the normalization condition $a_{i}^2+b_{i}^2=1, i=1, 2, \cdots, m$.

Assume that two observers ${\cal A}_i$ and ${\cal B}$ share $\ell_i$ generalized EPR states $|\Phi_{t_{i-1}+1}\rangle$, $|\Phi_{t_{i-1}+2}\rangle, \cdots, |\Phi_{t_{i}}\rangle$, where $t_{i}=\sum_{j=1}^i\ell_j$, $t_0=0$ and $\sum_{j=1}^k\ell_i\leq m$. We donot need to consider the entangled states owned by single observer because they can be locally prepared and measured.

Define the operators $\textbf{A}_{x_i}$ ($x_i=0,1$) on the particles of the observer ${\cal A}_i$ as
\begin{align*}
\textbf{A}_{x_i}=
\left\{
\begin{array}{ll}
\cos\theta_i \sigma_z^{\otimes \ell_i-1}\otimes {\bf I}_2+(-1)^{x_i}\sin\theta_i\sigma_x^{\otimes \ell_i},
 &\mbox{ for even } \ell_i;
\\
\cos\theta_i \sigma_z^{\otimes \ell_i}+(-1)^{x_i}\sin\theta_i\sigma_x^{\otimes \ell_i},
& \mbox{ for odd }\ell_i;
\end{array}
\right.
\tag{C2}
\end{align*}
where $\sigma_z$ and $\sigma_x$ are Pauli operators, ${\bf I}_2$ is the identity operator on one qubit system, $\textbf{X}^{\otimes l}$ denotes the $l$-fold tensor of the operator $\textbf{X}$, and $\theta_i\in [0, \frac{\pi}{2}]$, $i=1, 2, \cdots, k$.

Define the operators $\textbf{B}_{y}$ ($y=0,1$) on the particles of the observer ${\cal B}$ as
\begin{align*}
\textbf{B}_{y}=(\otimes _{i=1}^k {\textbf{B}}_{i,y})\otimes \textbf{B}_{r,y}
\tag{C3}
\end{align*}
where $\textbf{B}_{i,y}$ and $\textbf{B}_{r,y}$ are given by
\begin{align*}
\textbf{B}_{i,y}=
&
\left\{
\begin{array}{ll}
(1-y)\sigma_z^{\otimes \ell_i-1}\otimes {\bf I}_2+y \sigma_x^{\otimes \ell_i},
 &\mbox{ for even } \ell_i;
\\
(1-y)\sigma_z^{\otimes \ell_i}+y \sigma_x^{\otimes \ell_i},
 & \mbox{ for odd } \ell_i;
\end{array}
\right.
\tag{C4}
\end{align*}
and
\begin{align*}
\textbf{B}_{r,y}=(1-y)\sigma_z^{2m-2t_k}+y\sigma_x^{2m-2t_k}.
\tag{C5}
\end{align*}

Before continuing the proof, we need to prove that $\textbf{A}_{x_1}, \textbf{A}_{x_2}\cdots, \textbf{A}_{x_k}$ and $\textbf{B}_{y}$ are observables. Note that $\textbf{A}_{x_1}$, $\textbf{A}_{x_2}$, $\cdots, \textbf{A}_{x_k}$ and $\textbf{B}_y$ are performed on local systems, they satisfy the commutativity condition. Moreover, since $\textbf{A}_{x_1}$,  $\mathbf{A}_{x_2}$, $\cdots, \textbf{A}_{x_k}$ and $\textbf{B}_y$ are symmetric, it is sufficient to prove that they are positive semidefinite. In fact, we can prove that they are unitary Hermitian operators. For even $\ell_i$, from Eq.(C2) we obtain that
\begin{align*}
\textbf{A}_{x_i}^2=&{\bf I}_{2^{\ell_i}}+(-1)^{x_i}\sin\theta_i\cos\theta_i(\textbf{Y}^{\otimes \ell_i-1}\otimes \sigma_x
+(-\textbf{Y})^{\otimes \ell_i-1}\otimes \sigma_x)
\\
=&{\bf I}_{2^{\ell_i}},
\tag{C6}
\end{align*}
where the operator $\textbf{Y}$ is given by $\textbf{Y}=\sigma_z\sigma_x$, and ${\bf I}_{2^{\ell_i}}$ is the identity operator on the system of $\ell_i$ qubits. For odd $\ell_i$, from Eq.(C2) we obtain that
\begin{align*}
\textbf{A}_{x_i}^2=&{\bf I}_{2^{\ell_i}}+(-1)^{x_i}\sin\theta_i\cos\theta_i(\textbf{Y}^{\otimes \ell_i}+(-\textbf{Y})^{\otimes \ell_i})
\\
=&{\bf I}_{2^{\ell_i}}.
\tag{C7}
\end{align*}
Eqs.(C6) and (C7) imply that all the operators $\textbf{A}_{x_i}$s are unitary. Moreover, $\textbf{B}_y$ ($y=0,1$) are unitary Hermitian because all the operators $\textbf{B}_{i,y}$s and $\textbf{B}_{r,y}$ are products of Pauli operators and identity operator ${\bf I}_2$. In Eq.(C5), $\otimes _{i=1}^k \textbf{B}_{i,y}$ are measurement operators on the systems shared with the observers ${\cal A}_1$, ${\cal A}_2$, $\cdots$, ${\cal A}_k$; $\textbf{B}_{r,y}$ are measurement operators of the observer ${\cal B}$ on his own system that is not shared with other observers \cite{DD}.

Now, we continue the proof. From the equalities $\langle \Phi_i|\sigma_z\otimes \sigma_z|\Phi_i\rangle=1$, $\langle \Phi_i|\sigma_x\otimes \sigma_x|\Phi_i\rangle=2{a}_ib_i$, and $\langle \Phi_i|\sigma_z\otimes \sigma_x|\Phi_i\rangle$=$\langle \Phi_i|\sigma_x\otimes \sigma_z|\Phi_i\rangle=0$, we have $\frac{1}{2}\sum_{x_i=0}^1\langle\Xi_i|A_{x_i}\otimes B_{i,y=0}|\Xi_i\rangle=\cos\theta_i$, where $|\Xi_i\rangle=\otimes_{j=t_{i-1}+1}^{t_i}|\Phi_j\rangle$, $i=1, 2, \cdots, k$. So, from Eqs.(4), (5), and (C1)-(C5) we obtain that
\begin{align*}
I^q_{k+1,k}
=&\frac{1}{2^k}\sum_{x_1, x_2, \cdots, x_k=0,1}\langle \Xi|(\otimes_{i=1}^k\textbf{A}_{x_i})\otimes \textbf{B}_{y=0}|\Xi\rangle
\\
=&
\langle \Xi_r|\textbf{B}_{r,y=0}|\Xi_r\rangle
\prod_{i=1}^{k}(\frac{1}{2}\sum_{x_i=0}^1\langle\Xi_i|\textbf{A}_{x_i}\otimes \textbf{B}_{i,y=0}|\Xi_i\rangle)
\\
=&\prod_{i=1}^k\cos\theta_i,
\tag{C8}
\end{align*}
where we have taken use of the equality $\langle \Xi_r|\textbf{B}_{r,y=0}|\Xi_r\rangle=1$ with $|\Xi_r\rangle=\otimes_{j=m-t_k+1}^m|\Phi_j\rangle$.

Similarly, it is easy to get $\frac{1}{2}\sum_{x_i=0}^1(-1)^{x_i}\langle\Xi_i|\textbf{A}_{x_i}\otimes \textbf{B}_{i,y=1}|\Xi_i\rangle=2a_ib_i\sin\theta_i$, $i=1, 2, \cdots, k$. So, we can obtain that
\begin{align*}
J^q_{k+1,k}
=&\frac{1}{2^k}\sum_{x_1, x_2, \cdots, x_k=0,1}\!\!\!(-1)^{\sum_{i=1}^kx_i}\langle \Xi|(\otimes_{i=1}^k\textbf{A}_{x_i})
 \otimes \textbf{B}_{y=1}|\Xi\rangle
\\
=&
\langle \Xi_r|\textbf{B}_{r,y=1}|\Xi_r\rangle
\prod_{i=1}^{k}(\frac{1}{2}\sum_{x_i=0}^1(-1)^{x_i}\langle\Xi_i|\textbf{A}_{x_i}\otimes \textbf{B}_{i,y=1}|\Xi_i\rangle)
\\
=&\prod_{i=1}^k\prod_{j=1}^m\sin\theta_ic_j,
\tag{C9}
\end{align*}
where we have taken use of the equality $\langle \Xi_r|\textbf{B}_{r,y=1}|\Xi_r\rangle=\prod_{j=m-t_k+1}^mc_j$ with $c_j=2a_jb_j$.

From the presented Lemma in Appendix B1, Eqs.(C8) and (C9) imply that
\begin{align*}
\max_{\theta_1, \theta_2, \cdots, \theta_k}\{|I^q_{k+1,k}|^{\frac{1}{k}}+|J^q_{k+1,k}|^{\frac{1}{k}}\}
=&\sqrt{1+\prod_{i=1}^mc_i^{\frac{2}{k}}}
\\
>&1
\tag{C10}
\end{align*}
when all parameters $c_i$s satisfy $\prod_{i=1}^mc_i\not=0$, where the maximum is achieved when  $\cos\theta_i=1/\sqrt{1+\prod_{i=1}^mc_i^{2/k}}$ for all $i=1, \cdots, m$.

Note that all observables of the observer ${\cal B}$ are products of Pauli operators and identity operator. The expectation equals mathematically to that of the same operators being separately performed by all observers except for ${\cal A}_1, {\cal A}_2, \cdots, {\cal A}_k$ in the original network $\textsf{G}_q$ shown in Fig.1. Thus, $I^q_{k+1,k}$ and $J^q_{k+1,k}$ are essentially linear combinations of multi-partite quantum correlations generated by all observers of the network $\textsf{G}_q$. Combined with the inequality (C10), the multipartite quantum correlations of $\textsf{G}_q$ violate the nonlinear inequality (6). Consequently, there exist specific observables for each observer of the quantum network  $\textsf{G}_q$ with multiple independent observers such that the prediction of the quantum theory is inconsistent with the generalized local realism. $\square$

\subsection*{Appendix C2: Arbitrary bipartite entangled pure states as quantum resources}

We assume that the quantum resources consist of pure bipartite entangled pairs:
\begin{align*}
\hat{\Xi}=\otimes_{i=1}^{m}|\Phi_i\rangle,
\tag{C11}
\end{align*}
where $|{\Phi}_i\rangle=a_{i}|\phi_i\rangle|\psi_i\rangle
+b_i|\phi^\perp_i\rangle|\psi^\perp_i\rangle+\sum_{j\in {\cal I}_i}|\phi_{i,j}\rangle|\psi_{i,j}\rangle$ denote general bipartite entangled pure states with positive real coefficients $a_i, b_i$ satisfying $a_i^2+b_i^2\leq 1$ in Hilbert space $\mathbb{H}$, ${\cal I}_i$ is an index set satisfying that $\{|\phi_i\rangle, |\phi^\perp_i\rangle, |\phi_{i,j}\rangle, j\in {\cal I}_i\}$ is a set of orthogonal states in Hilbert space $\mathbb{H}_{i,1}$ and $\{|\psi_i\rangle, |\psi^\perp_i\rangle, |\psi_{i,j}\rangle, j\in {\cal I}_i\}$ is a set of orthogonal states in Hilbert space $\mathbb{H}_{i,2}$, and $\mathbb{H}_{i,1}\otimes \mathbb{H}_{i,2}=\mathbb{H}$, $i=1, 2, \cdots, m$.

Note that each bipartite entangled pure state can be decomposed into $|{\Phi}_i\rangle$ in Eq.(C11) with special orthogonal states $|\phi_i\rangle$, $|\phi^\perp_i\rangle$, $|\psi_i\rangle$, $|\psi^\perp_i\rangle$, $|\phi_{i,j}\rangle$, $|\psi_{i,j}\rangle$, $j\in {\cal I}_i$. $|{\Phi}_i\rangle$ is entangled if and only if $a_i\not=0$ and $b_i\not=0$ (up to permutations of basis states).

Assume that two observers ${\cal A}_i$ and ${\cal B}$ share $\ell_i$ bipartite entangled states $|{\Phi}_{t_{i-1}+1}\rangle$, $|{\Phi}_{t_{i-1}+2}\rangle$, $\cdots$, $|{\Phi}_{t_{i}}\rangle$ with $t_{i}=\sum_{j=1}^{i}\ell_j$ and $t_{0}=0$, $i=1, 2, \cdots, k$. Denote $\hat{\sigma}_{z,i}, \hat{\sigma}_{x,i}, \tilde{\sigma}_{z,i}, \tilde{\sigma}_{x,i}$ as the following matrices
\begin{align*}
\hat{\sigma}_{z,i}=&|\phi_i\rangle\langle \phi_i|-|\phi^\perp_i\rangle\langle \phi^\perp_i|,
\tag{C12}
\\
\hat{\sigma}_{x,i}=&|\phi_i\rangle\langle \phi^\perp_i|+|\phi^\perp_i\rangle\langle \phi_i|,
\tag{C13}
\\
\tilde{\sigma}_{z,i}=&|\psi_i\rangle\langle \psi_i|-|\psi^\perp_i\rangle\langle \psi^\perp_i|,
\tag{C14}
\\
\tilde{\sigma}_{x,i}=&|\psi_i\rangle\langle \psi^\perp_i|+|\psi^\perp_i\rangle\langle \psi_i|,
\tag{C15}
\end{align*}
where $i=1, 2, \cdots, m$.

Define the operators $\textbf{A}_{x_i}$ ($x_i=0,1$) on the system owned by the observer ${\cal A}_{i}$ as
\begin{align*}
\textbf{A}_{x_i}=
\left\{
\begin{array}{lll}
(\cos\theta_i (\otimes_{j=t_{i-1}+1}^{t_i-1}\hat{\sigma}_{z,j})\otimes \hat{\bf I}_{2,t_i}
\\
\quad +(-1)^{x_i}\sin\theta_i(\otimes_{j=t_{i-1}+1}^{t_i}\hat{\sigma}_{x,j}))\oplus \hat{\bf I}_{r,i},
&\mbox{ for even } \ell_i;
 \\
(\cos\theta_i (\otimes_{j=t_{i-1}+1}^{t_i}\hat{\sigma}_{z,j})
\\
\quad +(-1)^{x_i}\sin\theta_i (\otimes_{j=t_{i-1}+1}^{t_i}\hat{\sigma}_{x,j}))\oplus \hat{\bf I}_{r,i},
& \mbox{ for odd } \ell_i;
\end{array}
\right.
\tag{C16}
\end{align*}
where $\hat{\bf I}_{2,t_i}$ denotes the identity operator on the subspace spanned by $\{|\phi_{t_{i}}\rangle, |\phi^\perp_{t_i}\rangle\}$, $\hat{\mathbf{I}}_{r,i}$ denotes the identity operator on the orthogonal complement of the subspace $\otimes_{j=t_{i-1}+1}^{t_i}{\rm span}\{|\phi_j\rangle, |\phi^\perp_j\rangle\}$ in Hilbert space $\otimes_{j=t_{i-1}+1}^{t_i}\mathbb{H}_{j,1}$, span$\{|\phi_j\rangle, |\phi^\perp_j\rangle\}$ denotes the space by linearly superposing the representative vectors of $|\phi_j\rangle$ and $|\phi^\perp_j\rangle$, $\oplus$ denotes the direct sum of two operators performed on two orthogonal subspaces, and $\theta_i\in[0, \frac{\pi}{2}]$, $i=1, 2, \cdots, k$.

Define the operators $\textbf{B}_y$ ($y=0, 1$) on the systems owned by the observer ${\cal B}$ as
\begin{align*}
\textbf{B}_y=(\otimes _{i=1}^k \textbf{B}_{i,y})\otimes \textbf{B}_{r,y}
\tag{C17}
\end{align*}
where ${\textbf{B}}_{i,y}$ and $\textbf{B}_{r,y}$ are given by
\begin{align*}
\textbf{B}_{i,y}=
&
\left\{
\begin{array}{ll}
((1-y)(\otimes_{j=t_{i-1}+1}^{t_i-1}\tilde{\sigma}_{z,j})\otimes\tilde{\bf  I}_{2,t_i}
+y(\otimes_{j=t_{i-1}+1}^{t_i}\tilde{\sigma}_{x,j}))\oplus\tilde{\bf I}_{r,i},
& \mbox{ for even } \ell_i;
\\
((1-y)(\otimes_{j=t_{i-1}+1}^{t_i}\tilde{\sigma}_{z,j})
+y(\otimes_{j=t_{i-1}+1}^{t_i}\tilde{\sigma}_{x,j}))\oplus\tilde{\bf I}_{r,i},
& \mbox{ for odd } \ell_i;
\end{array}
\right.
\tag{C18}
\end{align*}
and
\begin{align*}
\textbf{B}_{r,y}=&
[(1-y)\otimes_{j=t_k+1}^{m} (\hat{\sigma}_{z,j}\otimes
\tilde{\sigma}_{z,j})
+y\otimes_{j=t_k+1}^{m}(\hat{\sigma}_{x,j}
\otimes \tilde{\sigma}_{x,j})]\oplus\tilde{\bf I}_{r}.
\tag{C19}
\end{align*}
Here, $\tilde{\bf I}_{2,t_i}$ denotes the identity operator on the subspace spanned by $|\psi_{t_{i}}\rangle$ and  $|\psi^\perp_{t_i}\rangle$, $\tilde{\mathbf{I}}_{r,i}$ denotes the identity operator on the orthogonal complement of the subspace $\otimes_{j=t_{i-1}+1}^{t_i}{\rm span}\{|\psi_j\rangle, |\psi^\perp_j\rangle\}$ in Hilbert space $\otimes_{j=t_{i-1}+1}^{t_i}\mathbb{H}_{j,2}$, and $\tilde{\bf I}_r$ denotes the identity operator on the orthogonal complement of the subspace $\otimes_{j=m-t_k+1}^{m}{\rm span}\{|\phi_j\rangle|\psi_j\rangle$, $|\phi^\perp_j\rangle|\psi^\perp_j\rangle\}$ in Hilbert space $\otimes_{j=m-t_{k}+1}^{m}\mathbb{H}_{j}$.

Similar to Eqs.(C6) and (C7), it is easy to prove that all operators $\textbf{A}_{x_i=0}$ and $\textbf{A}_{x_i=1}$ are unitary Hermitian, which can be used as the observables of the observers ${\cal A}_i$, $i=1, 2, \cdots, k$. Moreover, $\textbf{B}_{y=0}$ and $\textbf{B}_{y=1}$ can be used as the observables of the observer ${\cal B}$ because all operators $\textbf{B}_{i,1}, \textbf{B}_r$ are unitary Hermitian. Especially, in Eq.(C18), $\otimes _{i=1}^k \textbf{B}_{i,y}$ are measurement operators of the observer ${\cal B}$ on the systems shared with observers ${\cal A}_1, {\cal A}_2, \cdots, {\cal A}_k$. In Eq.(C19), $\textbf{B}_{r,y}$ are measurement operators of the observer ${\cal B}$ on his own systems that are not shared with other observers.

From Eqs.(C11)-(C15), we obtain that $\langle {\Phi}_i|\hat{\sigma}_{z,i}\otimes\tilde{\sigma}_{z,i}|{\Phi}_i\rangle
=a^2_i+b_i^2$, $\langle {\Phi}_i|\hat{\sigma}_{x,i}
\otimes\tilde{\sigma}_{x,i}|{\Phi}_i\rangle=2{a}_ib_i$, and $\langle {\Phi}_i|\hat{\sigma}_{z,i}
\otimes\tilde{\sigma}_{x,i}|{\Phi}_i\rangle$=$\langle {\Phi}_i|\hat{\sigma}_{x,i}\otimes\tilde{\sigma}_{z,i}|{\Phi}_i\rangle=0$.
Denote $|\hat{\Xi}_r\rangle=\otimes_{j=m-t_k+1}^m|{\Phi}_j\rangle$ and $|\hat{\Xi}_i\rangle=\otimes_{j=t_{i-1}+1}^{t_i}|{\Phi}_j\rangle$, where $i=1, 2, \cdots, k$. From Eqs.(C12)-(C19), we obtain the following equalities
\begin{align*}
&
\langle \hat{\Xi}_r|\textbf{B}_{r,y=0}|\hat{\Xi}_r\rangle=1,
\tag{C20}
\\
&\langle\hat{\Xi}_r|\textbf{B}_{r,y=1}|\hat{\Xi}_r\rangle
=1-\alpha+\beta,
\tag{C21}
\\
&\frac{1}{2}\sum_{x_i=0}^1\langle\hat{\Xi}_i|\textbf{A}_{x_i}\otimes \textbf{B}_{i,y=0}|\hat{\Xi}_i\rangle
=\alpha_i(\cos\theta_i-1)+1,
\tag{C22}
\\
&\frac{1}{2}\sum_{x_i=0}^1(-1)^{x_i}\langle\hat{\Xi}_i|\textbf{A}_{x_i}\otimes \textbf{B}_{i,y=1}|\hat{\Xi}_i\rangle=
\beta_i\sin\theta_i,
\tag{C23}
\end{align*}
where $\alpha=\prod_{j=t_k+1}^m(a_j^2+b_j^2)$, $\beta=\prod_{j=t_{k}+1}^{m}2a_jb_j$, $\alpha_i=\prod_{j=t_{i-1}+1}^{t_i}(a_j^2+b_j^2)$ and $\beta_i=\prod_{j=t_{i-1}+1}^{t_i}2a_jb_j$, $i=1, 2, \cdots, k$.

Now, from Eqs.(4), (5), and (C18)-(C22), we obtain that
 \begin{align*}
I^q_{k+1, k}=&\frac{1}{2^k}\sum_{x_1, x_2, \cdots, x_k}\langle \hat{\Xi}|(\otimes_{i=1}^k\textbf{A}_{x_i})\otimes \textbf{B}_{y=0}|\hat{\Xi}\rangle
\\
=&
\langle \hat{\Xi}_r|\textbf{B}_{r,y=0}|\hat{\Xi}_r\rangle
\prod_{i=1}^{k}(\frac{1}{2}\sum_{x_i=0}^1\langle\hat{\Xi}_i|\textbf{A}_{x_i}\otimes \textbf{B}_{i,y=0}|\hat{\Xi}_i\rangle)
\\
=&\prod_{i=1}^k((\cos\theta_i-1)\alpha_i+1).
\tag{C24}
\end{align*}

Similarly, from Eqs.(4), (5), and (C18)-(C23), we can obtain that
\begin{align*}
J^q_{k+1,k}=&\frac{1}{2^k}\sum_{x_1, x_2, \cdots, x_k}(-1)^{\sum_{i=1}^kx_i}\langle \Xi|(\otimes_{i=1}^k\textbf{A}_{x_i})\otimes \textbf{B}_{y=1}|\Xi\rangle
\\
=&
\langle \Xi_r|\textbf{B}_{r,y=1}|\Xi_r\rangle
\prod_{i=1}^{k}(\frac{1}{2}\sum_{x_i=0}^1(-1)^{x_i}\langle\Xi_i|\textbf{A}_{x_i}\otimes \textbf{B}_{i,y=1}|\Xi_i\rangle)
\nonumber\\
=&(1-\alpha+\beta)\prod_{i=1}^k\beta_i\sin\theta_i.
\tag{C25}
\end{align*}

Denote $\alpha_0=\max\{\alpha_1, \alpha_2, \cdots, \alpha_k\}$ and $\beta_0=\min\{\beta_1, \beta_2, \cdots, \beta_k\}$, where $0\leq \alpha_0, \beta_0\leq 1$. By setting $\theta_1=\theta_2=\cdots=\theta_k=\theta$ with $\cos\theta=\alpha_0/\sqrt{\alpha_0^2+\beta_0^2(1-\alpha+\beta)^2}$,  Eqs.(C24) and (C25) imply that
\begin{align*}
|I^q_{k+1,k}|^{\frac{1}{k}}+|J^q_{k+1,k}|^{\frac{1}{k}}
\geq &
\alpha_0\cos\theta+\beta_0(1-\alpha+\beta) \sin\theta-\alpha_0+1
\nonumber
\\
=&\sqrt{\alpha_0^2+\beta_0^2(1-\alpha+\beta)^2}-\alpha_0+1
\nonumber
\\
>&1
\tag{C26}
\end{align*}
when $\beta_0\not=0$ or $\alpha-\beta\not=1$, which are ensured by $\prod_{i=1}^ma_ib_i\not=0$.

Note that all observables of the observer ${\cal B}$ are product operators and direct sum of the identity operators. Hence, there exist observables for all observers except for ${\cal A}_1, {\cal A}_2, \cdots, {\cal A}_k$ of the original network $\textsf{G}_q$ shown in Fig.1 such that $I^q_{k+1,k}$ and $J^q_{k+1,k}$ are functions of the multipartite correlations crossing the whole network $\textsf{G}_q$. This completes the proof.

\subsection*{Appendix C3: Quantum resources consisting of bipartite entangled pure states and generalized GHZ states}

We assume that quantum resources consist of entangled pairs:
\begin{align*}
|\Theta\rangle=\otimes_{i=1}^{m_1}\otimes_{j=1}^{m_2}
|\Phi_i\rangle|\Psi_j\rangle,
\tag{C27}
\end{align*}
where $|\Phi_i\rangle$s are bipartite entangled pure states defined in Eq.(C11), and $|\Psi_j\rangle=\hat{a}_j|0\rangle^{\otimes s_j}+\hat{b}_j|1\rangle^{\otimes s_j}$ are generalized $s_j$-qubit GHZ states \cite{GHZ} with positive real coefficients $\hat{a}_j, \hat{b}_j$ satisfying $\hat{a}_j^2+\hat{b}_j^2\leq 1$, $j=1, 2, \cdots, m_2$.

Firstly, we assume that $s_1, s_2, \cdots, s_{m_1}$ are even integers. The observers ${\cal A}_i$ and ${\cal B}$ share $\ell_i$ bipartite entangled states $|{\Phi}_{t_{i-1}+1}\rangle$, $|{\Phi}_{t_{i-1}+2}\rangle$, $\cdots$, $|{\Phi}_{t_{i}}\rangle$, and $\hat{\ell}_i$ generalized GHZ states $|\Psi_{\hat{t}_{i-1}+1}\rangle$, $|\Psi_{\hat{t}_{i-1}+2}\rangle, \cdots$, $|\Psi_{\hat{t}_{i}}\rangle$, where $t_{i}=\sum_{j=1}^{i}\ell_j$, $\hat{t}_{i}=\sum_{j=1}^{i}\hat{\ell}_j$, and $t_{0}=\hat{t}_{0}=0$, $i=1, 2, \cdots, k$.

Using Eqs.(C12)-(C15), when $\ell_i\not=0$ we define the operators $\textbf{A}_{x_i}$ ($x_i=0,1$) on the system owned by the observer ${\cal A}_{i}$ as
\begin{align*}
\textbf{A}_{x_i}=
\left\{
\begin{array}{lll}
((\cos\theta_i(\otimes_{j=t_{i-1}+1}^{t_i-1}\hat{\sigma}_{z,j})\otimes \hat{\bf I}_2)\oplus\hat{\bf I}_{r,i})\otimes \sigma^{\otimes L_i}_{z}
\\
\quad +(-1)^{x_i}(\sin\theta_i
(\otimes_{j=t_{i-1}+1}^{t_i-1}\hat{\sigma}_{x,j})\oplus\hat{\bf I}_{r,i})\otimes \sigma^{\otimes L_i}_{x},
&  \mbox{ for even } K_i;
\\
(\cos\theta_i (\otimes_{j=t_{i-1}+1}^{t_i}\hat{\sigma}_{z,j})\oplus
\hat{\bf I}_{r,i})
\otimes\sigma^{\otimes L_i}_z
\\
\quad +(-1)^{x_i}(\sin\theta_i
(\otimes_{j=t_{i-1}+1}^{t_i-1}\hat{\sigma}_{x,j})\oplus
\hat{\bf I}_{r,i})\otimes \sigma^{\otimes L_i}_{x}),
& \mbox{ for odd } K_i;
\end{array}
\right.
\tag{C28}
\end{align*}
Otherwise, define $\textbf{A}_{x_i}$ as
\begin{align*}
\textbf{A}_{x_i}=
\left\{
\begin{array}{lll}
 \cos\theta_i(\sigma^{\otimes L_i-1}_{z}\otimes {\bf I}_2)
+(-1)^{x_i}\sin\theta_i \sigma^{\otimes L_i}_{x},
& \mbox{ for even } L_i;
\\
\cos\theta_i\sigma^{\otimes L_i}_{z}
+(-1)^{x_i}\sin\theta_i \sigma^{\otimes L_i}_{x},
& \mbox{ for odd } L_i;
\end{array}
\right.
\tag{C29}
\end{align*}
where $\hat{\bf I}_2, \hat{\bf I}_{r,i}$ are identity operators defined in Eq.(C16), ${\bf I}_2$ is the identity operator on single qubit, $L_i$ is the number of single quantum particles (belonging to the observer ${\cal A}_i$) in all generalized GHZ states shared with the observers ${\cal A}_i$ and ${\cal B}$, $K_i=\ell_i+L_i$, and $i=1, 2, \cdots, k$.

Using Eqs.(C12)-(C15) and (C17)-(C19), define the operators $\textbf{B}_y$ ($y=0,1$) on the system owned by the observer ${\cal B}$ as
\begin{align*}
\textbf{B}_y=(\otimes _{i=1}^k {\textbf{B}}_{i,y})\otimes {\textbf{B}}_{r,y},
\tag{C30}
\end{align*}
where ${\bf B}_{i,y}$ is given by
\begin{align*}
{\textbf{B}}_{i,y}=
&
\left\{
\begin{array}{ll}
(1-y)(((\otimes_{j=t_{i-1}+1}^{t_i-1}\tilde{\sigma}_{z,j})\otimes \hat{\bf I}_2)\oplus\tilde{\bf I}_{r,i})
\otimes \sigma^{\otimes N_i}_{z}
\\
\quad +y((\otimes_{j=t_{i-1}+1}^{t_i}\tilde{\sigma}_{x,j})\oplus\tilde{\bf I}_{r,j})
\otimes \sigma^{\otimes N_i}_{x},
& \mbox{ for even } K_i;
\\
(1-y)((\otimes_{j=t_{i-1}+1}^{t_i} \tilde{\sigma}_{z,j})\oplus\tilde{\bf I}_{r,i})
\otimes \sigma^{\otimes N_i}_{z}
\\
\quad +y((\otimes_{j=t_{i-1}+1}^{t_i}
\tilde{\sigma}_{x,j})\oplus\tilde{\bf I}_{r,i})\otimes \sigma^{\otimes N_j}_{x},
&  \mbox{ for odd } K_i;
\end{array}
\right.
\tag{C31}
\end{align*}
for $\ell_i\not=0$, or
\begin{align*}
{\textbf{B}}_{i,y}=
&
\left\{
\begin{array}{ll}
(1-y)(\sigma^{\otimes N_i-1}_{z}\otimes {\bf I}_2)
+y \sigma^{\otimes N_i}_{x},
& \mbox{ for even } N_i;
\\
(1-y)\sigma^{\otimes N_i}_{z}+y \sigma^{\otimes N_i}_{x},
& \mbox{ for odd } N_i;
\end{array}
\right.
\tag{C32}
\end{align*}
for $\ell_i=0$; and ${\textbf{B}}_{r,y}$ is given by
\begin{align*}
{\textbf{B}}_{r,y}=&(1-y)((\otimes_{j=t_k+1}^{m}\hat{\sigma}_{z,j}
\otimes
\tilde{\sigma}_{z,j})\oplus\hat{\bf I}_{r})\otimes\sigma^{\otimes N}_{z}
+y(\otimes_{j=t_k+1}^{m}(\hat{\sigma}_{x,j}
\otimes
\tilde{\sigma}_{x,j})\oplus\hat{\bf I}_{r})\otimes\sigma^{\otimes N}_{x}.
\tag{C33}
\end{align*}
Here, $N_i$ is the number of single particles (belonging to the observer ${\cal B}$) in all generalized GHZ states that are shared by the observers ${\cal B}$ and ${\cal A}_i$; and $N$ is the number of single particles in all generalized GHZ states (belonging to the observer ${\cal B}$) that are not shared by the observers ${\cal A}_1, {\cal A}_2, \cdots, {\cal A}_k$. $\tilde{\bf I}_{r,j}$ and $\tilde{\bf I}_r$ are identity operators defined in the respective Eq.(C18) and (C19).

Similar to Eqs.(C6) and (C7), we easily prove that $\textbf{A}_{x_i=0}$ and $\textbf{A}_{x_i=1}$ are unitary Hermitian, $i=1, 2, \cdots, k$. Thus, they can be used as the observables of the observers ${\cal A}_i$. Moreover, $\textbf{B}_{y=0}$ and $\textbf{B}_{y=1}$ can be used as the observable of the observer ${\cal B}$ because all the operators $\textbf{B}_{i,y}$s and $\textbf{B}_{r,y}$ are direct sum of generalized Pauli matrices defined in Eqs.(C12)-(C15) and the identity operators. In Eq.(C30), $\otimes _{i=1}^k\textbf{B}_{i,y}$ are measurement operators of the observer ${\cal B}$ on the particles shared with all the observers ${\cal A}_i$s while ${\textbf{B}}_{r,y}$ are measurement operators of the observer ${\cal B}$ on his own systems that are not  shared with other observers.

Denote $|\Theta_r\rangle=\otimes_{i=t_k+1}^{m_1}\otimes_{j=\hat{t}_k+1}^{m_2}
|{\Phi}_i\rangle|\Psi_j\rangle$ and $|\Theta_i\rangle=\otimes_{j=t_{i-1}+1}^{t_i}\otimes_{\jmath=\hat{t}_{i-1}+1}^{\hat{t}_i}
|{\Phi}_j\rangle|\Psi_\jmath\rangle$, where $i=1, 2, \cdots, k$. From Eqs.(C30)-(C33), we obtain the following equalities
\begin{align*}
&
\langle \Theta_r|\textbf{B}_{r,y=0}|\Theta_r\rangle=1,
\tag{C34}
\\
&\langle\Theta_r|\textbf{B}_{r,y=1}|\Theta_r\rangle
=1-\gamma+\delta,
\tag{C35}
\\
&\frac{1}{2}\sum_{x_i=0}^1\langle\Theta_i|\textbf{A}_{x_i}\otimes \textbf{B}_{i,y=0}|\Theta_i\rangle
=(\cos\theta_i-1)\gamma_i+1,
\tag{C36}
\\
&\frac{1}{2}\sum_{x_i=0}^1(-1)^{x_i}\langle\Theta_i|\textbf{A}_{x_i}\otimes \textbf{B}_{i,y=1}|\Theta_i\rangle=
\delta_i\sin\theta_i,
\tag{C37}
\end{align*}
where $\gamma=\prod_{j=t_k+1}^{m_1}(a_j^2+b_j^2)$,  $\delta=\prod_{j=t_{k}+1}^{m_1}\prod_{\jmath=\hat{t}_{k}+1}^{m_2}
4a_jb_j\hat{a}_\jmath \hat{b}_\jmath$, $\gamma_i=\prod_{j=t_{i-1}+1}^{t_i}(a_j^2+b_j^2)$, and $\delta_i=\prod_{j=t_{i-1}+1}^{t_i}\prod_{\jmath=\hat{t}_{i-1}+1}^{\hat{t}_i}$
$4a_jb_j\hat{a}_\jmath\hat{b}_\jmath$, $i=1, 2, \cdots, k$.

Now, from Eqs.(4), (5), (C27) and (C34)-(C37), we obtain that
 \begin{align*}
I^q_{k+1, k}=&\frac{1}{2^k}\sum_{x_1, x_2, \cdots, x_k}\langle \Theta|(\otimes_{i=1}^k\textbf{A}_{x_i})\otimes \textbf{B}_{y=0}|\Theta\rangle
\\
=&
\langle \Theta_r|\textbf{B}_{r,y=0}|\Theta_r\rangle
\prod_{i=1}^{k}(\frac{1}{2}\sum_{x_i=0}^1\langle\Theta_i|\textbf{A}_{x_i}\otimes \textbf{B}_{i,y=0}|\Theta_i\rangle)
\nonumber\\
=&\prod_{i=1}^k(\gamma_i(\cos\theta_i-1)+1)
\tag{C38}
\end{align*}
and
\begin{align*}
J^q_{k+1,k}=&\frac{1}{2^k}\sum_{x_1, x_2, \cdots, x_k}(-1)^{\sum_{i=1}^kx_i}\langle \Theta|(\otimes_{i=1}^k\textbf{A}_{x_i})\otimes \textbf{B}_{y=1}|\Theta\rangle
\\
=&
\langle \Theta_r|\textbf{B}_{r,y=1}|\Theta_r\rangle
\prod_{i=1}^{k}(\frac{1}{2}\sum_{x_i=0}^1(-1)^{x_i}\langle\Theta_i|\textbf{A}_{x_i}\otimes \textbf{B}_{i,y=1}|\Theta_i\rangle)
\nonumber\\
=&
(1-\gamma+\delta)\prod_{i=1}^k\delta_i\sin\theta_i.
\tag{C39}
\end{align*}

Denote $\gamma_0=\max\{\gamma_1, \gamma_2, \cdots, \gamma_k\}$ and $\delta_0=\min\{\delta_1, \delta_2, \cdots, \delta_k\}$, where $0\leq \gamma_0, \delta_0\leq 1$. Note that $\gamma\geq \delta$ and $\gamma_i\geq \delta_i$ , $i=1, 2, \cdots, k$. By setting $\theta_1=\theta_2=\cdots =\theta_k=\theta$ with $\cos\theta=\gamma_0/\sqrt{\gamma_0^2+\delta_0^2(1-\gamma+\delta)^2}$, Eqs.(C38) and (C39) imply that
\begin{align*}
|I^q_{k+1,k}|^{\frac{1}{k}}+|J^q_{k+1,k}|^{\frac{1}{k}}
\geq &
\gamma_0\cos\theta+\delta_0(1-\gamma+\delta) \sin\theta-\gamma_0+1
\nonumber
\\
=&\sqrt{\gamma_0^2+\delta_0^2(1-\gamma+\delta)^2}-\gamma_0+1
\nonumber
\\
>&1
\tag{C40}
\end{align*}
when $\delta_0\not=0$ or $\gamma-\delta\not=1$, which are ensured by $\prod_{i=1}^{m_1}\prod_{j=1}^{m_2}a_ib_i\hat{a}_j\hat{b}_j\not=0$.

Note that all the observables of the observer ${\cal B}$ are product operators and direct sum of the identity operators. Thus, there exist observables for all the observers except for ${\cal A}_1, {\cal A}_2, \cdots, {\cal A}_k$ in the network $\textsf{G}_q$ shown in Fig.1 such that $I^q_{k+1,k}$ and $J^q_{k+1,k}$ are functions of multipartite quantum correlations crossing the whole network $\textsf{G}_q$. This completes the proof.

Now, for general integers $s_j\geq 3$, assume that $s_1, s_2, \cdots, s_{d}$ are odd integers and $s_{d+1}, s_{d+2}, \cdots, s_{m_{2}}$ are even integers. The main idea is to replace one Pauli matrix $\sigma_z$ with ${\bf I}_2$ for each generalized GHZ state with odd number of particles. Note that all observables of the observer ${\cal B}$ are product operators and direct sum of the identity operators. Similar to Eqs.(C30)-(C33), we can easily redefine $\textbf{B}_{y}$ by replacing one Pauli operator $\sigma_z$ with ${\bf I}_2$ (In experiment, one can perform a measurement under the basis $\{|\pm\rangle\}$, and then output $1$) for each generalized GHZ state with odd number of particles. It implies that some observer may use commutative operators of $\{{\bf I}_2, \sigma_z\}$, which can be regarded as classical outcomes. All the operators $\textbf{A}_{x_i}$s are unchanged. Thus, we can obtain the same quantities of $I^q_{k+1,k}$ and $J^q_{k+1,k}$ in respective Eq.(C38) and (C39). This  is derived from the equalities $\langle \Psi_j|\sigma_z^{\otimes s_j-1}\otimes{\bf I}_2|\Psi_j\rangle=1$,  $\langle \Psi_j|\sigma_x^{\otimes s_j}|\Psi_j\rangle=2\hat{a}_j\hat{b}_j$ and $\langle \Psi_j|\sigma_z^{\otimes t}\otimes\sigma_x^{\otimes s_j-t}|\Psi_j\rangle=0$ with $0<t<s_j$, for odd integer $s_j$. The followed proof is omitted.

\section*{Appendix C4: The maximal violation of Theorem A}

In this subsection we prove the possibility of the maximal violation with respect to Tsirelson's bound presented in Eq.(7). It is sufficient to consider the inequality (C40) for general quantum resources. In fact, the proof of the maximal violation is equivalent to $\max_{\theta_i}\{|I_{k+1,k}|^{\frac{1}{k}}+|J_{k+1,k}|^{\frac{1}{k}}\}=\sqrt{2}$. From Eqs.(C38) and (C39) and $\delta_i\leq \gamma_i$, we obtain that
\begin{align*}
|I^q_{k+1,k}|^{\frac{1}{k}}+|J^q_{k+1,k}|^{\frac{1}{k}}
\leq &|\prod_{i=1}^k(\gamma_i(\cos\theta_i-1)+1)|^{\frac{1}{k}}
+|\prod_{i=1}^k\delta_i|^{\frac{1}{k}}
|\prod_{i=1}^k\sin\theta_i|^{\frac{1}{k}}
\tag{C41}
\\
=&|\prod_{i=1}^k\cos\theta'_i|^{\frac{1}{k}}
+|\prod_{i=1}^k\delta_i|^{\frac{1}{k}}|\prod_{i=1}^k
\sin\theta_i|^{\frac{1}{k}}
\tag{C42}
\\
\leq &\cos(\frac{1}{k}\sum_{i=1}^k\theta'_i)
+\prod_{i=1}^k\delta_i^{\frac{1}{k}}\sin(\frac{1}{k}\sum_{i=1}^k\theta_i),
\tag{C43}
\end{align*}
where $\cos\theta_i':=\gamma_i(\cos\theta_i-1)+1$ with $\theta_i'\in [0, \pi/2]$ in Eq.(C42), and the inequality (C43) is from the presented Lemma in Appendix B1. The equality in Eq.(C43) holds when $\theta_1'=\theta_2'=\cdots=\theta_k'$ (which is denoted as $\theta'$ for simplicity) and $\theta_1=\theta_2=\cdots=\theta_k$ (which is denoted as $\theta$). These conditions imply that $\gamma_1=\gamma_2=\cdots=\gamma_k$, which is denoted as $\gamma$. Thus, we obtain that
\begin{align*}
\max_{\theta}\{|I^q_{k+1,k}|^{\frac{1}{k}}+|J^q_{k+1,k}|^{\frac{1}{k}}\}
\leq &
\gamma(\cos\theta-1)+1
+\prod_{i=1}^k\delta_i^{\frac{1}{k}}\sin\theta
\\
\leq&\sqrt{\gamma^2+\prod_{i=1}^k\delta_i^{\frac{2}{k}}}+1-\gamma
\tag{C44}\\
\leq &(\sqrt{2}-1)\gamma+1
\tag{C45}\\
\leq& \sqrt{2},
\tag{C46}
\end{align*}
where the inequality (C44) is from the inequality $x\sin\theta+y\cos\theta\leq \sqrt{x^2+y^2}$; the inequality (C45) is from the inequality $\prod_{i=1}^k\delta_i^{\frac{2}{k}}\leq \prod_{i=1}^k\gamma_i^{\frac{2}{k}}
=\gamma^2$; and the inequality (C46) is from the inequality $\gamma\leq 1$.

Note that the equality in Eq.(C41) holds when $\gamma=\delta$, which follows that $|{\Phi}_{t_{k}+1}\rangle$, $|{\Phi}_{t_{k}+2}\rangle, \cdots, |{\Phi}_{m_1}\rangle$ are maximally entangled EPR states, and $|{\Phi}_{\hat{t}_{k}+1}\rangle, |{\Phi}_{\hat{t}_{k}+2}\rangle, \cdots, |{\Phi}_{m_2}\rangle$ are maximally entangled GHZ states. The equality in Eq.(C44) holds when $\cos\theta=\gamma/\sqrt{\gamma^2+\prod_{i=1}^k\gamma_i^{2/k}}$. The equality in Eq.(C45) holds when $a_i=b_i$ for all $i$. The equality in Eq.(C46) holds when $\gamma=1$ which implies $a_i=b_i=1/\sqrt{2}$ for all $i$. Consequently, the violation in Eq.(C40) is maximal with respect to Tsirelson's bound in Eq.(7) when quantum resources consist of the maximally entangled EPR states and GHZ states.

\section*{Appendix C5: Quantum resources consisting of unknown generalized EPR states and GHZ states}

Consider the situation that all the parameters of generalized EPR states and generalized GHZ states are unknown or partially unknown for some observers. For example, the maximally entangled EPR state evolves to a partially entangled state because of a non-isolated system. This problem has not been theoretically considered in terms of the nonlocality. Fortunately, Eqs.(C38)-(C40) allow us to probabilistically complete the task of verifying violation. By setting $\min\{\theta_1, \theta_2, \cdots, \theta_k\}=\theta$, from Eqs.(C38) and (C39), we obtain that
\begin{align*}
|I^q_{k+1,k}|^{\frac{1}{k}}+|J^q_{k+1,k}|^{\frac{1}{k}}
\geq &
\gamma_0\cos\theta+\delta_0(1-\gamma+\delta) \sin\theta-\gamma_0+1
\nonumber
\\
\approx&\gamma_0(1-\frac{1}{2}\theta^2)+\delta_0(1-\gamma+\delta)\theta -\gamma_0+1
\nonumber
\\
=&\theta(\delta_0(1-\gamma+\delta)-\frac{1}{2}\gamma_0\theta) +1
\nonumber
\\
>&1
\tag{C47}
\end{align*}
when $\delta_0(1-\gamma+\delta)>\frac{1}{2}\gamma_0\theta$, which is ensured by $\theta<2\delta_0(1-\gamma+\delta)$ and $\delta-\gamma<1$. It implies a simple method for each observer who chooses an observable with a small $\theta_i>0$ for unknown EPR states and GHZ states as quantum resources.

\section*{Appendix D: Proof of Theorem B}

In this section, we prove Theorem B. For convenience, we take use of the notations defined in Appendix C. Assume that a quantum network with $n$ observers has an equivalent network shown in Fig.2, i.e, there are $k$ independent observers who do not share quantum resources. In what follows, we assume that noisy quantum sources consist of Werner states:
\begin{align*}
\rho=&(\otimes_{i=1}^{m_1}
(v_i|\Phi_i\rangle\langle \Phi_i|+\frac{1-v_i}{4}\mathbbm{1}_4))
\otimes (\otimes_{j=1}^{m_2}(w_j|\Psi_j\rangle\langle \Psi_j|+\frac{1-w_j}{2^{s_j}}\mathbbm{1}_{2^{s_j}})),
\tag{D1}
\end{align*}
where $|\Phi_i\rangle$ are generalized EPR states defined in Eq.(C1), and $|\Psi_j\rangle$ are generalized GHZ states defined in Eq.(C27). $\mathbbm{1}_{2^{s_j}}$ are the $2^{s_j}$ square identity matrices.

Denote the subsystems $\rho_{0}, \rho_{i}$ as
\begin{align}
\rho_{0}=&(\otimes_{i=t_k+1}^{m_1}(v_i|\Phi_i\rangle\langle \Phi_i|+\frac{1-v_i}{4}\mathbbm{1}_4))\otimes(\otimes_{j=\hat{t}_k+1}^{m_2}(w_j|\Psi_j\rangle\langle \Psi_j|+\frac{1-w_j}{2^{s_j}}\mathbbm{1}_{2^{s_j}})),
\tag{D2}
\\
\rho_{i}=&(\otimes_{j=t_{i-1}+1}^{t_i}(v_i|\Phi_i\rangle\langle \Phi_i|+\frac{1-v_i}{4}\mathbbm{1}_4))\otimes(\otimes_{\jmath=\hat{t}_{i-1}+1}^{\hat{t}_i}(w_j|\Psi_j\rangle\langle \Psi_j|+\frac{1-w_j}{2^{s_j}}\mathbbm{1}_{2^{s_j}})),
\tag{D3}
\end{align}
where $i=1, 2, \cdots, k$.

From Eqs.(C30)-(C33) (without $\hat{\mathbf{I}}_{r,i}$ and $\tilde{\mathbf{I}}_{r,i}$), we obtain the following equalities
\begin{align*}
&
{\rm Tr}(\textbf{B}_{r,y=0}\rho_{0} )=\prod_{i=t_k+1}^{m_1}\prod_{j=\hat{t}_k+1}^{m_2}v_iw_j,
\tag{D3}
\\
&{\rm Tr}(\textbf{B}_{r,y=1}\rho_{0})=\prod_{i=t_k+1}^{m_1}\prod_{j=\hat{t}_k+1}^{m_2}v_ic_iw_j\hat{c}_j,
\tag{D4}
\\
&\frac{1}{2}\sum_{x_i=0}^1{\rm Tr}( (\textbf{A}_{x_i}\otimes \textbf{B}_{i,y=0})\rho_{i})
=\prod_{j=t_{i-1}+1}^{t_i}\prod_{\jmath=\hat{t}_{i-1}+1}^{\hat{t}_i}\cos\theta_j v_jw_\jmath,
\tag{D5}
\\
&\frac{1}{2}\sum_{x_i=0}^1(-1)^{x_i}{\rm Tr}( (\textbf{A}_{x_i}\otimes \textbf{B}_{i,y=1})\rho_{i})=
\prod_{j=t_{i-1}+1}^{t_i}\prod_{\jmath=\hat{t}_{i-1}+1}^{\hat{t}_i}\sin\theta_j v_jc_jw_\jmath\hat{c}_\jmath,
\tag{D6}
\end{align*}
where $c_j=2a_jb_j$ and $\hat{c}_\jmath=2\hat{a}_\jmath \hat{b}_\jmath$.

From Eqs.(D1)-(D6), we get that
\begin{align*}
I^q_{k+1,k}=&\sum_{x_1, x_2, \cdots, x_k=0,1}{\rm Tr}[((\otimes_{i=1}^k\textbf{A}_{x_i})\otimes \textbf{B}_{y=0})\rho]
\\
=&
{\rm Tr}(\textbf{B}_{r,y=0}\rho_{0})
\prod_{i=1}^{k}(\frac{1}{2}\sum_{x_i=0}^1{\rm Tr}( (\textbf{A}_{x_i}\otimes \textbf{B}_{i,y=0})\rho_{i}))
\\
=&\prod_{i=1}^k\prod_{j=1}^{m_1}\prod_{\jmath=1}^{m_2}\cos\theta_iv_jw_\jmath
\tag{D7}
\end{align*}
and
 \begin{align*}
J^q_{k+1,k}=&\sum_{x_1, x_2, \cdots, x_k=0,1}(-1)^{\sum_{i=1}^kx_i}{\rm Tr}(((\otimes_{i=1}^k\textbf{A}_{x_i})\otimes \textbf{B}_{y=1} )\rho)
\\
=&
{\rm Tr}( \textbf{B}_{r,y=1}\rho_{0})
\prod_{i=1}^{k}(\frac{1}{2}\sum_{x_i=0}^1(-1)^{x_i}
{\rm Tr}( (\textbf{A}_{x_i}\otimes \textbf{B}_{i,y=1})\rho_{i}))
\\
=&\prod_{i=1}^k\prod_{j=1}^{m_1}
\prod_{\jmath=1}^{m_2}\sin\theta_ic_jv_j\hat{c}_\jmath w_\jmath.
\tag{D8}
\end{align*}

From the presented Lemma in Appendix B1, Eqs.(D7) and (D8) imply that
  \begin{align*}
\max_{\theta_1, \theta_2, \cdots, \theta_k}\{|I^q_{k+1,k}|^{\frac{1}{k}}+|J^q_{k+1,k}|^{\frac{1}{k}}\}
=&\prod_{i=1}^{m_1}\prod_{j=1}^{m_2}v_i^{\frac{1}{k}}w_j^{\frac{1}{k}}\sqrt{1+\prod_{i=1}^{m_1}\prod_{j=1}^{m_2}c_i^{\frac{2}{k}}
\hat{c}_j^{\frac{2}{k}}},
\tag{D9}
\end{align*}
where the maximum is achieved when $\cos\theta_1=\cos\theta_2=\cdots =\cos\theta_k=1/\sqrt{1+\prod_{i=1}^{m_1}\prod_{j=1}^{m_2}(c_i\hat{c}_j)^{\frac{2}{k}}}$. Eq.(D9) implies that the product of critical viabilities $v_j^*, \hat{v}^*_\jmath$ is given by
\begin{align*}
\prod_{i=1}^{m_1}\prod_{j=1}^{m_2}v^*_iw^*_j
\leq \frac{1}{(1+\prod_{i=1}^{m_1}\prod_{j=1}^{m_2}c_i^{\frac{2}{k}}
\hat{c}_j^{\frac{2}{k}})^{\frac{k}{2}}}
\tag{D10}
\end{align*}
for which the multipartite quantum correlations violate the Bell inequality presented in Eq.(6).

\section*{Appendix E: The comparison of the visibilities for the network shown in Fig.3(a)}

For the chain-shaped network shown in Fig.3(a), assume that the total system consists of Werner states:
\begin{align*}
\rho=\otimes_{i=1}^{n-1}\rho_{\Phi_i},
\tag{E1}
\end{align*}
where $\rho_{\Phi_i}=v_i|\Phi_i\rangle\langle \Phi_i|+\frac{1-v_i}{4}\mathbbm{1}_4$ with generalized EPR states $|\Phi_i\rangle$ and $4$ square identity matrix $\mathbbm{1}_4$, and $0<v_i<1$, $i=1, 2, \cdots, n-1$. Here, the observers ${\cal A}_{i}$ and ${\cal A}_{i+1}$ share a generalized EPR state in the form of $\rho_{\Phi_i}$, $i=1, 2, \cdots, n-1$.

Note that there are at most $k=\lceil \frac{n}{2}\rceil$ independent observers who do not share entangled states, where $\lceil x\rceil$ denotes the smallest integer no less than $x$. From Theorem B, the upper bound of the product of critical visibilities $v_i^*$ is given by
\begin{align*}
\frac{1}{(\max_{\theta}\{|I^q_{n,k}|^{\frac{1}{k}}+|J^q_{n,k}|^{\frac{1}{k}}\})^{k}}
=&\frac{1}{(1+\prod_{i=1}^{n-1}c_i^{\frac{2}{k}})^{\frac{k}{2}}}
\\
= &\frac{1}{\sqrt{\sum_{i=0}^{k}f_i(c)}}
\\
\geq &\frac{1}{\sum_{j=0}^{n-1}g_j(c)}
\tag{E2}
\\
\geq &\hat{v}_i
\tag{E3}
\\
=&
\prod_{j=1}^{n-1}\frac{1}{\sqrt{1+c_j^{2}}},
\tag{E4}
\end{align*}
for which the multipartite correlations of this quantum network violate the Bell inequality presented in Eq.(6), where $f_i(c)=(^{k}_i)c^{2i/k}$, $g_j(c)=(^n_j)c^{2j/n}$, $(^{t}_s)$ is a binomial coefficient given by $(t(t-1)\cdots (t-s+1))/(s(s-1)\cdots 1)$, $c=\prod_{j=1}^{n-1}c_j$, and $c_j=2a_jb_j$, $i=0, 1, \cdots, k$; $j=0, 1, \cdots, n-1$. Here, the inequality (E2) is from the inequalities $f_i(c)\leq g_i(c)$ which are derived from $(^n_i)\geq (^{k}_i)$ and $c^{2i/n}>c^{2i/k}$ with $c\leq 1$, $i=0, 1, \cdots, k$; the inequality (E3) is from the algebraic inequality $\sum_{{\cal J}_i}\prod_{j\in {\cal J}_i}c_{j}^2\geq (^{n}_i) c^{2i/n}$, where ${\cal J}_i$ denotes the subset of $\{0, 1, \cdots, n-1\}$ with $i$ integers; and the summation is evaluated over all possible subsets ${\cal J}_i$, $i=0, 1, \cdots, n-1$. $\hat{v}_i$ in Eq.(E3) denotes the known upper bound of the visibility of EPR state $|\Phi_i\rangle$ \cite{GH}.

\begin{figure}
\begin{center}
\resizebox{265pt}{235pt}{\includegraphics{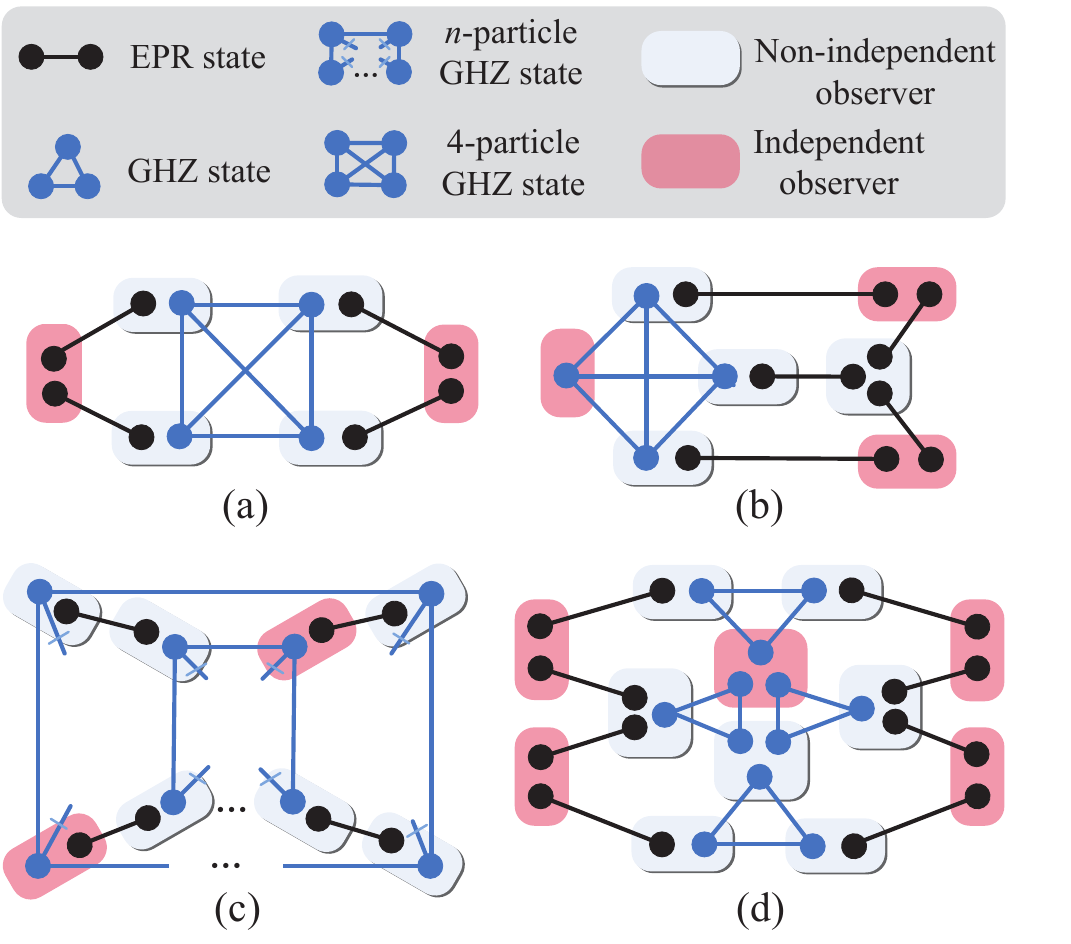}}
\end{center}
\caption{\small (Color online) Simple networks whose multipartite correlations violate the nonlinear Bell inequalities presented in Eq.(6). (a) Cyclic network of two loops consisting of one four-partite generalized GHZ state and 4 generalized EPR states. (b) Butterfly network consisting of one four-partite generalized GHZ state and 5 generalized EPR states. (c) Hybrid network consisting of two $n$-partite generalized GHZ states and $n$ generalized EPR states. Here $n\geq 2$. (d) Boat network consisting of 4 generalized GHZ states and 8 generalized EPR states. Here, each dot denotes one particle. Each $n$-partite generalized GHZ state is represented with a complete graph of $n$ vertexes.}
\label{fig-S3}
\end{figure}

\section*{Appendix F: Supplementary networks}

In this section, we provide some additional networks shown in Fig.S3 going beyond the presented examples in main text. Specially, there are two loops in the first network shown in Fig.S3(a), where quantum resources consist of one four-partite generalized GHZ state and 4 generalized EPR states. Two red squares shown in the Figure represent independent observers. From Theorem A, the multipartite quantum correlations violate the nonlinear Bell inequality presented in Eq.(6) with $k=2$. The second example is a butterfly network shown in Fig.S3(b), where quantum resources consist of one four-partite generalized GHZ state and 5 generalized EPR states. It is interesting in classical networks \cite{ACL} or quantum networks for multicast task \cite{Hay}. The multipartite quantum correlations violate the nonlinear Bell inequality presented in Eq.(6) with $k=3$, where three red squares represent independent observers. There are multiple loops in the third network shown in Fig.S3(c), where quantum resources consist of  two $n$-partite generalized GHZ states and $n$ generalized EPR states. Each of $2n$ observers has two particles. Two red squares represent independent observers. The last one is a boat-type network shown in Fig.S3(d), which consist of 4 generalized GHZ states and 8 generalized EPR states. The multipartite correlations of the network violate the nonlinear Bell inequality presented in Eq.(6) with $k=5$. In addition to these examples, one can easily construct lots of networks depending on special tasks.

\section*{Appendix G: The non-multilocality of quantum network consisting of general noisy states}

In this section, we provide some sufficient conditions of the non-multilocality for quantum networks consisting of general noisy states. For convenience, denote $\sigma_1:={\bf I}_2$, $\sigma_2:=\sigma_x$, $\sigma_3:=\sigma_y$, and $\sigma_4:=\sigma_z$.

Assume that a quantum network has an equivalent network shown in Fig.2. Consider the noisy states:
\begin{align*}
\rho=&\otimes_{i=1}^{m_1}\otimes_{j=1}^{m_2}\hat{\rho}_{i}\tilde{\rho}_j,
\tag{G1}
\end{align*}
where $\hat{\rho}_{i}$ are two-particle systems in the state
$\hat{\rho}_i=\frac{1}{4}\sum_{j_1,j_2=1}^4v^i_{j_1j_2}\sigma_{j_1}\otimes \sigma_{j_2}$ with $v^i_{11}=1$, and $\tilde{\rho}_j$ are $s_j$-particle systems in the state $\tilde{\rho}_j=\frac{1}{2^{s_j}}\sum_{i_1,i_2, \cdots, i_{s_j}=1}^4w^{j}_{i_1i_2\cdots i_{s_j}}\otimes_{t=1}^{s_j}\sigma_{i_{t}}
$ with $w^{j}_{11\cdots 1}=1$. It has been proved that $|v^i_{j_1j_2}|\leq 1$ for every quantum states of two qubits \cite{HHH}. For any quantum state of $s_j$ qubits, we obtain that $w^{j}_{i_1i_2\cdots i_{s_j}}\leq 1$ from ${\rm Tr}(\tilde{\rho}^2_j)\leq 1$.

Here, we take use of the notations defined in Appendix C3. Denote the subsystems $\rho_{\Theta_r}$, $\rho_{\Theta_i}$ as
\begin{align}
\rho_{0}=&\otimes_{i=t_k+1}^{m_1}\otimes_{j=\hat{t}_k+1}^{m_2}
\hat{\rho}_{i}\otimes\tilde{\rho}_j,
\tag{G2}
\\
\rho_{i}=&\otimes_{j=t_{i-1}+1}^{t_i}\otimes_{\jmath=\hat{t}_{i-1}+1}^{\hat{t}_i}
\hat{\rho}_{j}\otimes\tilde{\rho}_\jmath,
\tag{G3}
\end{align}
where $i=1, 2, \cdots, k$.

From Eqs.(C30)-(C33) (without $\hat{\mathbf{I}}_{r,i}$ and $\tilde{\mathbf{I}}_{r,i}$) and Eqs.(G2)-(G3), we obtain the following results
\begin{align*}
&
{\rm Tr}(\textbf{B}_{r,y=0}\rho_{0} )=\prod_{i=t_k+1}^{m_1}\prod_{j=\hat{t}_k+1}^{m_2}v^i_{44}w^{j}_{44\cdots 4},
\tag{G4}
\\
&{\rm Tr}(\textbf{B}_{r,y=1}\rho_{0})
=\prod_{i=t_k+1}^{m_1}\prod_{j=\hat{t}_k+1}^{m_2}v^i_{22}w^{j}_{22\cdots 2},
\tag{G5}
\\
&\left|\frac{1}{2}\sum_{x_i=0}^1{\rm Tr}( (\textbf{A}_{x_i}\otimes \textbf{B}_{i,y=0})\rho_{i})\right|\geq \prod_{j=t_{i-1}+1}^{t_i}\prod_{\jmath=\hat{t}_{i-1}+1}^{\hat{t}_i}\cos\theta_j |v^j_{44}w^{\jmath}_{44\cdots 4}|,
\tag{G6}
\\
&\left|\frac{1}{2}\sum_{x_i=0}^1(-1)^{x_i}{\rm Tr}( (\textbf{A}_{x_i}\otimes \textbf{B}_{i,y=1})\rho_{i})\right|\geq
\prod_{j=t_{i-1}+1}^{t_i}\prod_{\jmath=\hat{t}_{i-1}+1}^{\hat{t}_i}\sin\theta_j |v^j_{22}w^{\jmath}_{22\cdots 2}|,
\tag{G7}
\end{align*}
where Eqs.(G6) and (G7) are derived from the inequalities $v^{j}_{11}=1\geq v^{j}_{44}$ and $ w^{\jmath}_{11\cdots 1}=1\geq w^{\jmath}_{44\cdots 4}$ when $K_i$ is even and $\ell_i\not=0$; or $L_i$ is even and $\ell_i=0$.

From Eqs.(G1)-(G9), we obtain that
\begin{align*}
|I^q_{k+1,k}|=&\left|\sum_{x_1, x_2, \cdots, x_k}{\rm Tr}(((\otimes_{i=1}^k\textbf{A}_{x_i})\otimes \textbf{B}_{y=0})\rho
)\right|
\\
=&
|{\rm Tr}( \textbf{B}_{r,y=0}\rho_{0})|
\prod_{i=1}^{k}\left|\frac{1}{2}\sum_{x_i=0}^1{\rm Tr}( (\textbf{A}_{x_i}\otimes \textbf{B}_{i,y=0})\rho_{i})\right|
\\
\geq &\prod_{i=1}^k\prod_{j=1}^{m_1}\prod_{\jmath=1}^{m_2}\cos\theta_iv^j_{44}
w^\jmath_{44\cdots 4}
\tag{G8}
\end{align*}
and
 \begin{align*}
|J^q_{k+1,k}|=&|\sum_{x_1, x_2, \cdots, x_k}(-1)^{\sum_{i=1}^kx_i}{\rm Tr}(((\otimes_{i=1}^k\textbf{A}_{x_i})\otimes \textbf{B}_{y=1} )\rho)|
\\
=&
|{\rm Tr}( \textbf{B}_{r,y=1}\rho_{0})|
\prod_{i=1}^{k}\left|\frac{1}{2}\sum_{x_i=0}^1(-1)^{x_i}
{\rm Tr}( (\textbf{A}_{x_i}\otimes \textbf{B}_{i,y=1})\rho_{i})\right|
\\
\geq&\prod_{i=1}^k\prod_{j=1}^{m_1}
\prod_{\jmath=1}^{m_2}\sin\theta_iv^j_{22}
w^\jmath_{22\cdots 2}.
\tag{G9}
\end{align*}

From the presented Lemma in Appendix B1, Eqs.(G8) and (G9) imply that
  \begin{align*}
\max_{\theta_1, \theta_2, \cdots, \theta_k}\{|I^q_{k+1,k}|^{\frac{1}{k}}+|J^q_{k+1,k}|^{\frac{1}{k}}\}
\geq&\sqrt{\prod_{i=1}^{m_1}\prod_{j=1}^{m_2}(v^i_{22}
w^j_{22\cdots 2})^{\frac{2}{k}}+\prod_{i=1}^{m_1}\prod_{j=1}^{m_2}(v^i_{44}
w^j_{44\cdots 4})^{\frac{2}{k}}},
\tag{G10}
\end{align*}
where the maximum is achieved when $\cos\theta_1=\cos\theta_2=\cdots =\cos\theta_k=\prod_{i=1}^{m_1}\prod_{j=1}^{m_2}(v^i_{22}
w^j_{22\cdots 2})^{\frac{1}{k}}$\\
 /$\sqrt{\prod_{i=1}^{m_1}\prod_{j=1}^{m_2}(v^i_{22}
w^j_{22\cdots 2})^{\frac{2}{k}}+\prod_{i=1}^{m_1}\prod_{j=1}^{m_2}(v^i_{44}
w^j_{44\cdots 4})^{\frac{2}{k}}}$.

Eq.(G10) implies a sufficient condition
  \begin{align*}
\prod_{i=1}^{m_1}\prod_{j=1}^{m_2}(v^i_{22}
w^j_{22\cdots 2})^{\frac{2}{k}}+\prod_{i=1}^{m_1}\prod_{j=1}^{m_2}(v^i_{44}
w^j_{44\cdots 4})^{\frac{2}{k}}
> 1
\tag{G11}
\end{align*}
for which that the multipartite quantum correlations violate the Bell inequality presented in Eq.(6). A simple sufficient condition is given by
\begin{align*}
v^i_{22}, v^i_{44},  w^j_{22\cdots 2}, w^j_{44\cdots 4}\geq \frac{\sqrt{2}}{2}
\tag{G12}
\end{align*}
for all $i=1, 2, \cdots, m_1$ and $j=1, 2, \cdots, m_2$.

\begin{figure}
\begin{center}
\resizebox{180pt}{205pt}{\includegraphics{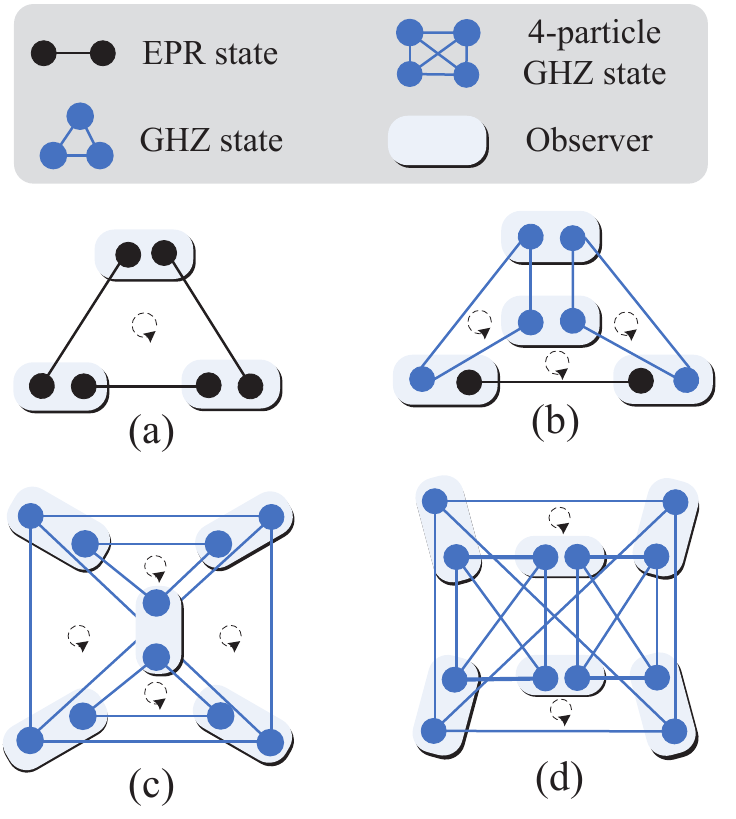}}
\end{center}
\caption{\small (Color online) Simple networks cannot be characterized with the presented Bell inequalities. (a) Triangle cyclic network consisting of 3 EPR states. (b) Special cyclic network consisting of 2 GHZ states and one EPR state. (c) Symmetric cyclic network consisting of multi-partite GHZ states. (d) Door-type network consisting of 3 four-partite GHZ states.  Each colored square denotes one observer of a network.}
\label{fig-S4}
\end{figure}

Moreover, if $v^i_{22}=v^i_{44}=w^j_{44\cdots 4}=w^j_{22\cdots 2}=1$, the violation is maximal with respect to the bound presented in Eq.(7). Note that the condition in Eq.(G11) is independent of all the coefficients except for $v^i_{22}, v^i_{44}, w^j_{44\cdots 4}, w^j_{22\cdots 2}$. This property is useful in applications.

\section*{Appendix H: Inefficient networks}

Here, we provide some simple networks that cannot be characterized with the nonlinear Bell inequalities presented in Eq.(6). The first one is the cyclic network shown in Fig.S4(a) consisting of EPR states. This network is also different from the triangle cyclic network consisting of 2 GHZ states. For the second network shown in Fig.S4(b), there are three cyclic subnetworks, where quantum resources consist of 2 GHZ states and one EPR state. There are 4 cyclic subnetworks in the network shown in Fig.S4(c), where quantum resources consist of one four-partite GHZ state and 2 GHZ states. The last one is a door-type network shown in Fig.S4(d) with 4 cyclic subnetworks, where quantum resources consist of 3 four-partite GHZ states. For these simple cyclic networks, there does not exist $k\geq2$ independent observers who do not share entangled states. Hence, it is interesting to explore new Bell inequalities for these special networks. Actually, we conjecture the linear inequalities should be useful for these networks.


\begin{thebibliography}{99}
\bibitem{DD}In these definitions of observables $A_{x_i}, B_y$, there exist no shared EPR state that is measured with an identity operator by two parties. For an even $\ell_i$, if the particle of the $t_j$-th EPR states of the observer ${\cal B}_{y}$ belongs to the observer ${\cal C}$ in the  network $G_q$, ${\cal C}$ should have other particles; Otherwise, ${\cal C}$ has only one particle. By replacing the observer ${\cal A}_i$ with the observe ${\cal C}$, it follows a new equivalent network with at least $k$ independent observers, where $\ell_i=1$.
\end{thebibliography}
\end{document}